\documentclass[twocolumn,pre,superscriptaddress,showpacs,preprintnumbers,amsmath,amssymb,floatfix]{revtex4-1}

\usepackage{graphicx}
\usepackage[usenames,dvipsnames]{color}
\usepackage{amsmath}

\newcommand{\ve}[1]{\mathbf{#1}}

\newlength{\onefig}
\begin{document}

\title{Ultrasoft primitive model of polyionic solutions: structure, aggregation, and dynamics}

\author{Daniele Coslovich} 
\affiliation{Laboratoire Charles Coulomb UMR 5221, Universit{\'e}
  Montpellier 2 and CNRS, Montpellier, France}

\author{Jean-Pierre Hansen} 
\affiliation{Department of Chemistry, University of Cambridge,
  Cambridge, United Kingdom \\ and \\ PECSA, Universit{\'e} Pierre et Marie
  Curie, Paris, France}

\author{Gerhard~Kahl} 
\affiliation{Institut f\"ur Theoretische Physik and Center for
  Computational Materials Science (CMS), Technische Universit\"at
  Wien, Wien, Austria}
\thanks{Copyright (2011) American Institute of Physics. This article
  may be downloaded for personal use only. Any other use requires
  prior permission of the author and the American Institute of
  Physics.}

\date{\today}

\begin{abstract}
We introduce an ultrasoft core model of interpenetrating polycations and polyanions with continuous Gaussian charge distributions, to investigate polyelectrolyte aggregation in dilute and semi-dilute, salt-free solutions. The model is studied by a combination of approximate theories (random phase approximation and hypernetted chain theory) and numerical simulations. The calculated pair structure, thermodynamics, phase diagram and polyion dynamics of the symmetric version of the model (the ``ultrasoft restricted primitive model'' or UPRM) differ from the corresponding properties of the widely studied ``restricted primitive model'' (RPM) where ions have hard cores. At sufficiently low temperatures and densities, oppositely charged polyions form weakly interacting, polarizable neutral pairs. The clustering probabilities, dielectric behavior and electrical conductivity point to a line of sharp conductor-insulator transitions in the density-temperature plane. At very low temperatures the conductor-insulator transition line terminates near the top of a first order coexistence curve separating a high-density, liquid phase from a low-density, vapor phase. The simulation data hint at a tricritical behavior, reminiscent of that observed of the two-dimensional Coulomb Gas, which contrasts with the Ising criticality of its three-dimensional counterpart, the RPM.
\end{abstract}


\maketitle

\section{Introduction}
\label{sec:introduction}

Ever since the pioneering work of Gouy~\cite{Gouy__1910}, Chapman
\cite{chapman__1913} and of Debye and H\"uckel~\cite{debye__1923},
electrostatic interactions are known to play a dominant role in
determining the structure, dynamics and phase behavior of ionic
liquids and solutions, 
as well as in governing the colloid stability of polyelectrolyte
solutions, complex biomolecular assemblies, and related soft matter
systems. In the case of solutions and melts of microscopic cations and
anions (microions), like Na$^+$ and Cl$^-$, a widely studied model
system is the primitive model (PM) of oppositely charged hard spheres,
which is now known to undergo a phase separation into a very dilute
phase of mostly paired ions (``Bjerrum pairs''~\cite{bjerrum__1926})
and a more concentrated solution of non-aggregated ions, for
sufficiently strong Coulomb coupling~\cite{stell_critical_1976}; for a
review of simulation work, see~\cite{panagiotopoulos_simulations_2005}.

Moving to the mesoscopic scale, charged colloidal particles
(macroions) are stabilized in aqueous dispersions by the formation of
electric double-layers of microscopic co- and counterions, leading to
a screened Coulombic repulsion between equally charged, ``dressed''
colloids; for an overview see~\cite{hansen__2000}. Despite this
purely repulsive effective interaction between colloids, so-called
``volume terms'', associated with the self energy of individual
electric double-layers, induce a phase separation between dilute and
concentrated colloidal dispersions~\cite{van_roij_phase_1999}, which
may be regarded as the highly asymmetric counterpart of the
fluid-fluid transition of the PM [5, 6]. Recently the experimental and
theoretical attention has shifted to the rich phase behavior of such
``colloidal electrolytes'', where the polyanions and polycations are
highly charged, hard colloidal~\cite{leunissen_ionic_2005} or
nanometric
particles~\cite{ryden_monte_2005,dahirel_ion-mediated_2009}, usually
in the presence of added
salt~\cite{hynninen_cuau_2006,caballero_complete_2007,sanz_gel_2008}.
These mesoscopic electrolytes may thus be regarded as a
generalization of the PM, where the pure Coulombic interactions are
replaced by screened repulsive and attractive electrostatic (or
Yukawa) forces.

Another important class of complex ionic systems are synthetic or
natural polyelectrolytes, i.e., solutions of charged polymer chains
\cite{barrat__1995}, where overall charge neutrality is ensured by
either microscopic or mesoscopic counterions. The latter case
corresponds to a binary system of polymeric polyanions and
polycations, which have been shown to aggregate into neutral or
charged polyelectrolyte complexes (complex coacervation), in the
presence or absence of added salt
\cite{philipp__1989,tsuchida__1994,dautzenberg_polyelectrolyte_1997,buchhammer_formation_2003}.
Since polyions are now flexible, worm-like charged objects, they
cannot be reasonably modeled by charged hard spheres (like their
colloidal counterparts), unless they collapse into quasi-spherical
globules like certain folded proteins (e.g. lysozyme). Modern
theoretical descriptions of polyelectrolyte complexation are usually
based on statistical field-theoretic formulations
\cite{edwards__1965,fredrickson_equilibrium_2006}, within a
perturbative~\cite{castelnovo__2001} or a simulation
\cite{lee_complex_2008} framework.

In the present paper we introduce and investigate a simple model of
polyanion/polycation aggregation in polyelectrolyte solutions without
added salt. Swollen polymer coils in good solvent are known to
interpenetrate easily. In fact the free energy penalty for two
self-avoiding polymer coils to fully overlap, such that their
centers-of-mass (CM) coincide, is of the order of twice the thermal
energy $k_{\rm B}T$, independently of molecular weight
\cite{grosberg__1982,dautenhahn_monte_1994,bolhuis_accurate_2001}. A
convenient coarse-grained representation of dilute and semi-dilute
polymer coils in good solvent reduces the coils to ultrasoft,
interpenetrating particles; the effective pair potentials between CM's
of interpenetrating coils can be extracted from full monomer Monte
Carlo (MC) simulations of the underlying microscopic model by a
systematic inversion procedure
\cite{dautenhahn_monte_1994,bolhuis_accurate_2001}. The resulting pair
potential is well represented by a Gaussian of amplitude $\simeq 2
k_{\rm B}T$, and width of the order of the radius of gyration, $R_g$,
of the polymer coils
\cite{dautenhahn_monte_1994,bolhuis_accurate_2001}. Here we generalize
the ultrasoft core representation to globally neutral binary solutions
of polyanions and polycations in a dielectric continuum representing
the solvent. The total charge of each polyion is assumed to be
smeared over the volume of the coil according to a quenched
Gaussian distribution of width $R_g$, centered on the coil CM, ensuring
that the total electrostatic interaction between two coils of equal or
opposite signs remains finite, even at full overlap. In particular the
``Coulomb collapse'' of oppositely charged ions, which is prevented by
the presence of the hard core in the PM, is bypassed in the present
model of oppositely charged polyelectrolytes by averaging over the
spatial extension of the polyion charge distributions.

In the following sections we investigate the pair structure,
thermodynamics and polyion dynamics of this ``ultrasoft primitive
model'' (UPM) by a combination of approximation schemes borrowed from
the theory of ionic liquids, and of Monte Carlo (MC) and Molecular
Dynamics (MD) simulations. Special emphasis is laid on the static and
dynamic characterization of the polyion aggregation and complexation,
which are expected to induce a conductor-insulator (CI) transition at
low temperatures and concentrations and eventually to phase separation
as in the case of the PM
\cite{stell_critical_1976,panagiotopoulos_simulations_2005,levin_criticality_1996,valeriani_ion_2010}.
The complete phase diagram of the symmetric version of the UPM (the
``restricted'' UPM or URPM), which differs considerably from that of
the RPM, will be presented in a subsequent paper. A preliminary account
of parts of this work was published elsewhere~\cite{coslovich_clustering_2011}.

\section{The model}
\label{sec:model}

We consider a binary system of $N_+$ polycations of charge $Q_+ = Z_+
e$ and $N_-$ polyanions of charge $Q_- = Z_- e$ (where $e$ is the
proton charge), moving in a dielectric continuum of dielectric
permittivity $\epsilon '$ (the ``solvent'' in its ``primitive''
representation) and confined to a volume $V$. If $n_\alpha =
N_\alpha/V$ ($\alpha = +, -$) are the corresponding number densities,
overall charge neutrality requires that:
\begin{equation}
Z_+ n_+ + Z_- n_- = 0
\end{equation}
while the total polyion number density is $n = n_+ + n_-$. Since the
system under consideration is supposed to be a model for
polyelectrolyte coils, the polyions are not point particles, but their
charges are smeared over a volume of the order of the cube of
their radius of gyration $(R_{g+}$ or $R_{g-}$), according to a
Gaussian charge distribution centered on the position ${\bf r}_i$ of
the CM of each polyion ($1 \le i \le N = N_+ + N_-$). If $r$ is the
distance relative to that position, the normalized charge distribution
in units of $e$ of a polyion is assumed to be $Z_\alpha
\rho_\alpha(r)$ where
\begin{equation}
\rho_\alpha(r) = 
\left( \frac{1}{2 \pi \sigma_\alpha^2} \right)^{3/2}
\exp[-r^2/2 \sigma_\alpha^2]
\label{eq:charge_density}
\end{equation}
with $\sigma_\alpha$ of the order of $R_{g \alpha}$ and $\alpha = +, -$;
its Fourier transform is
\begin{equation}
\hat \rho_\alpha(k) = 
\int {\rm e}^{i {\bf k}{\bf r}} \rho_\alpha({\bf r}) d {\bf r} 
= \exp[- k^2 \sigma^2/2] .
\label{rho_k}
\end{equation}

The electrostatic potential $\varphi_\alpha(r)$ generated by the above
charge distribution obeys Poisson's equation (in esu)
\begin{equation} 
\nabla^2 \varphi_\alpha(r) = 
- \frac{4 \pi Z_\alpha e}{\epsilon '} \rho_\alpha (r) .
\end{equation}
Taking Fourier transforms of both sides, and remembering Eq.~(\ref{rho_k}), one finds
\begin{equation}
\hat \varphi_\alpha(k) = 
\frac{4 \pi Z_\alpha e}{\epsilon ' k^2} \exp[-k^2 \sigma_\alpha^2/2] .
\end{equation}
Inverse Fourier transformation leads to
\begin{equation}
\varphi_\alpha(r) = 
\frac{Z_\alpha e}{\epsilon ' r} {\rm erf} 
\left( r/ \sqrt{2} \sigma_\alpha \right)
\end{equation}
where ${\rm erf}(x)$ is the error function. In the point particle
limit, $\sigma_\alpha \to 0$, $\varphi_\alpha(r)$ reduces to $Z_\alpha
e/\epsilon ' r$, as expected. For finite $\sigma_\alpha$,
$\varphi_\alpha(r)$ remains finite as $r \to 0$.

The pair potential between an $\alpha$-polyion and a $\beta$-polyion
at a CM-CM distance $r$ is given by
\begin{equation}
v_{\alpha \beta}(r) = 
\int \varphi_\alpha(r') Z_\beta e \rho_\beta(| {\bf r} - {\bf r}'|) d {\bf r} ' .
\end{equation}
Applying the convolution theorem, the Fourier transform is
\begin{equation}
\hat v_{\alpha \beta} (k) = 
Z_\beta e \hat \varphi_\alpha(k) \hat \rho_\beta(k) = 
\frac{4 \pi Z_\alpha Z_\beta e^2}{\epsilon' k^2}
\exp[-k^2 \sigma_{\alpha \beta}^2]
\end{equation}
where $\sigma_{\alpha \beta}^2 = (\sigma_\alpha^2 + \sigma_\beta^2)/2$.

Inverse Fourier transformation finally yields the set of three pair
potentials between identical or opposite polyions
\begin{equation}
v_{\alpha \beta}(r) = \frac{Q_\alpha Q_\beta}{\epsilon ' r} 
{\rm erf} \left( r/2 \sigma_{\alpha \beta} \right) .
\label{eq:potential}
\end{equation}
This pair potential remains finite at full overlap, i.e., as $r \to 0$
\begin{equation}
v_{\alpha \beta}(r) \underset{r \to 0}{\sim}
u_{\alpha \beta} \left[ 1 - \frac{r^2}{12 \sigma_{\alpha \beta}^2}
+ \frac{r^4}{160 \sigma_{\alpha \beta}^4} - {\cal O}(r^6) \right]
\label{eq:v_origin}
\end{equation}
where the overlap energies are
\begin{equation}
u_{\alpha \beta} = 
\frac{Q_\alpha Q_\beta}{\sqrt{\pi} \epsilon' \sigma_{\alpha \beta}} .
\end{equation}

At large distances $v_{\alpha \beta}(r)$ goes over to the Coulombic
pair potential between point ions:
\begin{equation}
v_{\alpha \beta}(r) \underset{r \to \infty}{\sim}
\frac{Q_\alpha Q_\beta}{\epsilon ' r} 
\left[ 1 - \frac{2 \sigma_{\alpha \beta}}{\sqrt{\pi} r} 
{\rm e}^{- r^2/\sigma_{\alpha \beta}^2} \right]
\end{equation}

In order to introduce reduced (dimensionless) physical quantities, we
define the following convenient length, energy and time scales
\begin{subequations}
\begin{align}
{\rm length~scale:}\quad & \sigma = \sigma_{+-} \\ 
{\rm energy~scale:}\quad &  
u = - u_{+-} = \frac{Q_+ | Q_-|}{\sqrt{\pi} \epsilon ' \sigma_{+-}} \\ 
{\rm time~scale:}\quad &  
\tau = \left( \frac{m \sigma_{+-}^2}{u} \right)^{1/2}
\end{align}
\end{subequations}
where $m$ is the smaller of the polyanion and polycation masses $m_-$
and $m_+$. Reduced distances, densities, times and temperatures are
defined as
$$
r^* = r/\sigma ~~~~~~ n_\alpha^* = n_\alpha \sigma ^3 ~~~~~~  
t^* = t / \tau  ~~~~~~ T^* = k_{\rm B} T/u
$$
and in the following only reduced units will be used and asterisks are
dropped for convenience. Another, related dimensionless parameter is
the reduced Bjerrum length for monovalent ions, $l_{\rm B}$, which is
identical to the Coulomb coupling parameter $\Gamma$ commonly used in
the characterization of strongly coupled plasmas (see e.g.
\cite{baus_statistical_1980})
\begin{equation}
\Gamma = l_{\rm B} = \frac{e^2}{\epsilon ' k_{\rm B} T \sigma} \,.
\end{equation}
$\Gamma$ is directly related to the reduced inverse temperature
$\beta$ for a system of polyions by
\begin{equation}
\Gamma = \frac{\sqrt{\pi}}{| Z_+ Z_-|} \beta =  \frac{\sqrt{\pi}}{| Z_+ Z_-|} \frac{1}{T} \,.
\end{equation}

Two relevant special cases of the UPM are the following:
\begin{itemize}
\item[(a)] The restricted model (URPM) is the symmetric version where
  polyions are of the same size ($\sigma_+ = \sigma_- = \sigma$) and
  of opposite charge ($Q_+ = - Q_- = Q$). This is the model introduced
  in our preliminary communication~\cite{coslovich_clustering_2011},
  and most of the results in the subsequent sections will be for the
  URPM. Note that the set of reduced units introduced above is
  different from the one we employed in Ref.~\cite{coslovich_clustering_2011}.
\item[(b)] A fully asymmetric version of the model (UAPM) consists of
  polyanions ($Z =|Z_-| \gg 1$) and microscopic cations ($Z_+ = 1$),
  with $\sigma_- \gg \sigma_+$. This version provides a simple
  representation of an anionic polyelectrolyte in good solvent and its
  counterions.
\end{itemize}

At sufficiently low temperature and density, polyions and polycations
of the UPM are expected to cluster into neutral or charged aggregates,
as in the case of the PM
\cite{bjerrum__1926,stillinger_ion-pair_1968,gillan_liquid-vapour_1983,levin_criticality_1996}.
In the $T \to 0$ limit, the ground state of the symmetric version (the
URPM) is achieved by associating the $N$ polyions into $N/2$ neutral,
non-interacting pairs of total energy
\begin{equation}
U_0 = - \frac{N}{2} u .
\label{eq:u_0}
\end{equation}
This energy is finite and extensive, so that the URPM is
thermodynamically stable according to Ruelle's stability criterion
\cite{ruelle_statistical_1999}.

Similar considerations are expected to apply to asymmetric versions of
the UPM, but the ground state analysis is much less straightforward,
except when $| Z_-|$ is an integer multiple of $Z_+$ (or
conversely). To simplify notations we consider the case where $Z_+ =
1$ and $| Z_-| = Z$, corresponding to the UAPM. One polyanion and $Z$
counterions may then collapse into a neutral point cluster and the
corresponding ground-state energy of $N_+/Z$ neutral clusters would
be
\begin{eqnarray} \nonumber
U_0 & =  &
\frac{N_+}{Z} \left[ Z v_{+-}(r=0) + \frac{Z(Z-1)}{2} v_{++}(r=0) \right] =
\\ 
& = & - N_+ u \left[1 - \frac{Z-1}{Z} \frac{1}{2^{3/2}} 
\left( 1 + \Sigma^2 \right)^{1/2} \right]
\end{eqnarray}
where $\Sigma = \sigma_- / \sigma_+ \gg 1$.

For the ground state to be stable, $U_0$ must be negative. Assuming
$Z$ to scale like $\Sigma^3$ (volume charge), $U_0$ is negative
provided $Z$ satisfies the inequality
$$
Z^2 > \frac{(Z-1)^2}{8} \left( 1 + Z^{2/3} \right)
$$
which requires $Z \alt 22$. For larger anion to cation charge ratios the
neutral clusters are unstable, and the ground state can no longer be a
system of neutral point clusters.

Before presenting the results of our calculations for the UPM in the
following sections, two remarks are in order. First, the UPM is a
purely Coulombic system and no other force fields are involved. In
particular, we do not include the ultrasoft entropic repulsion between
interpenetrating coils mentioned in the Introduction. Apart from
obvious reasons of simplicity, this is physically justified at low
temperatures (where most of the interesting physics will be shown to
occur), where Coulombic interactions dominate the effective entropic
interactions, which scale like $T$. Secondly, a model somewhat similar
to the UPM was used previously to investigate a very different
Coulombic system namely a semi-classical Hydrogen plasma under
astrophysical conditions of high temperatures and densities
\cite{hansen_microscopic_1981}.

\section{Pair structure and thermodynamics}
\label{sec:structure_thermo}

The local pair structure of the UPM is characterized by the three
partial pair distribution functions $g_{++}(r)$, $g_{+-}(r)$, and
$g_{--}(r)$; the corresponding pair correlation functions $h_{\alpha
  \beta}(r) = g_{\alpha \beta}(r) -1$ go to zero at long distances in
the disordered fluid phases. Their Fourier transforms $\hat h_{\alpha
  \beta}(k)$ are related to those of the direct correlation functions
$\hat c_{\alpha \beta}(k)$ by the set of three coupled
Ornstein-Zernike (OZ) relations~\cite{hansen_theory_2006}; solving the
latter for the $\hat h_{\alpha \beta}(k)$, and introducing the
concentrations $x_\alpha = n_\alpha/n$, we find
\begin{subequations}
\begin{align}
\hat h_{++}(k) & = \frac{1}{D(k)} 
\left[ \hat c_{++}(k) - x_- \Delta(k) \right] 
\label{eq:def_h_pp} 
\\ 
\hat h_{+-}(k) & \equiv \hat h_{-+}(k) = \frac{\hat c_{+-}(k)}{D(k)} 
\label{eq:def_h_pm}
\\ 
\hat h_{--}(k) & = \frac{1}{D(k)} 
\left[ \hat c_{--}(k) - x_+ \Delta(k) \right] 
\label{eq:def_h_mm}
\end{align}
\end{subequations}
where
\begin{eqnarray} \nonumber
\Delta(k) & = & \hat c_{++}(k) \hat c_{--}(k) - \hat c_{+-}^2(k) \\ 
\nonumber
D(k) & = & \left[ 1 - x_+ c_{++}(k)\right] \left[ 1 - x_- c_{--}(k) \right]\\
 & & - x_+ x_- \hat c_{+-}^2(k)
\label{eq:def_delta_d}
\end{eqnarray}
and all Fourier transforms are dimensionless, i.e., $\hat f(k) = n
\int \exp(i {\bf k}{\bf r}) f(r) d {\bf r}$.

The partial structure factors $S_{\alpha \beta}(k)$, which are
measurable by X-ray or neutron diffraction experiments, are directly
related to the $\hat h_{\alpha \beta}(k)$ through
\cite{hansen_theory_2006}
\begin{equation}
S_{\alpha \beta}(k) = 
\frac{1}{N} \langle \rho_{\bf k}^\alpha \rho_{- {\bf k}}^\beta \rangle 
= x_\alpha \delta_{\alpha \beta} + x_\alpha x_\beta \hat h_{\alpha \beta}(k)
\end{equation}
where the Fourier components of the local density operators are
\begin{equation}
\rho_{\bf k}^\alpha = \sum_{i=1}^{N_\alpha} \exp(i {\bf k}\cdot {\bf r}_{i \alpha}) .
\end{equation}

It is convenient to introduce the linear combinations corresponding to
total number ($N$) and charge ($C$) densities
\begin{subequations}
\begin{align}
\rho_{\bf k}^N & = \rho_{\bf k}^+ + \rho_{\bf k}^- \\
\rho_{\bf k}^C & = Z_+ \rho_{\bf k}^+ + Z_- \rho_{\bf k}^- 
\end{align}
\end{subequations}
and the related structure factors
\begin{subequations}
\begin{align}
S_{NN}(k) & = \frac{1}{N} \langle \rho_{\bf k}^N \rho_{-{\bf k}}^N \rangle 
= \sum_\alpha \sum_\beta S_{\alpha \beta}(k) 
\label{eq:def_s_nn} \\
S_{NC}(k) & = \frac{1}{N} \langle \rho_{\bf k}^N \rho_{-{\bf k}}^C \rangle 
= \sum_\alpha \sum_\beta Z_\beta S_{\alpha \beta}(k) 
\label{eq:def_s_nc} \\
S_{CC}(k) & = \frac{1}{N} \langle \rho_{\bf k}^C \rho_{-{\bf k}}^C \rangle 
= \sum_\alpha \sum_\beta Z_\alpha Z_\beta S_{\alpha \beta}(k) \,.
\label{eq:def_s_cc}
\end{align}
\end{subequations}
The charge-charge structure factor obeys the Stillinger-Lovett limit,
valid for a conducting medium~\cite{stillinger_ion-pair_1968}
\begin{equation}
\lim_{k \to 0} \frac{\kappa_{\rm D}^2}{k^2} \frac{S_{CC}(k)}{\bar Z^2} 
= 1 
\label{eq:stillinger_lovett_limit}
\end{equation}
where $\bar Z^2 = x_+ Z_+^2 + x_- Z_-^2$ and $\kappa_{\rm D}$ is the
inverse Debye screening length
\begin{equation}
\kappa_{\rm D}^2 = \kappa_{{\rm D}+}^2 + \kappa_{{\rm D}-}^2 = 
4 \pi n_+ Z_+^2 l_{\rm B} + 4 \pi n_- Z_-^2 l_{\rm B} . 
\label{eq:kappa_d}
\end{equation}

Knowledge of the pair structure gives access to a number of
thermodynamic properties via standard relations
\cite{hansen_theory_2006}. Using the dimensionless variables defined
in Section \ref{sec:model}, the reduced excess internal energy per
polyion is given by
\begin{subequations}
\begin{align} \nonumber
u^{\rm ex} & = \frac{\beta U^{\rm ex}}{N} \\ \nonumber 
& = 2 \pi n \Gamma \int_0^\infty 
\left \{ x_+^2 Z_+^2 h_{++}(r) {\rm erf}(r/2 \sigma_+) \right.\\ \nonumber 
&  \left. + 2 x_+ x_- Z_+ Z_- h_{+-}(r) {\rm erf}(r/2 \sigma_{+-}) \right.  \\ 
&  \left. + x_-^2 Z_-^2 h_{--}(r) {\rm erf}(r/2 \sigma_-) \right \} r dr 
\label{eq:u_ex_r} \\ \nonumber
& = \frac{\Gamma}{\pi}
\int_0^\infty \left \{ x_+^2 Z_+^2 \hat h_{++} (k) {\rm e}^{-k^2 \sigma_+^2} \right.\\ \nonumber
& \left. + 2x_+x_- Z_+Z_- \hat h_{+-} (k) {\rm e}^{-k^2 \sigma_{+-}^2} \right.\\ 
& \left. + x_-^2 Z_-^2 \hat h_{--} (k) {\rm e}^{-k^2 \sigma_-^2} \right \} dk 
\label{eq:u_ex_k} 
\end{align}
\end{subequations}
where the transition from the first to the second of the above
equations has been achieved by using Parseval's theorem. 
The
wavenumbers $k$ in Eq.~(\ref{eq:u_ex_k}) are dimensionless,
i.e., expressed in units of $\sigma^{-1} \equiv \sigma_{+-}^{-1}$.

Similarly, the dimensionless equation of state (with $P$ the
pressure) is given by:
\begin{subequations}
\begin{align} \nonumber
\frac{\beta P}{n} & = 1 + \frac{u^{\rm ex}}{3}  \\ \nonumber
& - \frac{2 \Gamma \sqrt{\pi}}{3} n \int_0^\infty \left \{
\frac{x_+^2 Z_+^2}{\sigma_+} h_{++}(r) {\rm e}^{-r^2/4 \sigma_+^2} \right. \\ \nonumber 
& \left. + \frac{2x_+x_- Z_+Z_-}{\sigma_{+-}} h_{+-}(r) {\rm e}^{-r^2/4 \sigma_{+-}^2} \right. \\  
& \left. + \frac{x_-^2 Z_-^2}{\sigma_-} h_{--}(r) {\rm e}^{-r^2/4 \sigma_-^2} \right \}
r^2 dr = 
\label{eq:p_r} \\ \nonumber
& = 1 + \frac{u^{\rm ex}}{3}  \\ \nonumber
&  - \frac{2 \Gamma}{3 \pi} \int_0^\infty \left \{
x_+^2 Z_+^2 \sigma_+^2 \hat h_{++}(k) {\rm e}^{-k^2 \sigma_+^2} \right.\\\nonumber 
& \left. + 2x_+x_- Z_+Z_- \sigma_{+-}^2 \hat h_{+-}(k) {\rm e}^{-k^2 \sigma_{+-}^2} 
\right . \\ 
& \left. +x_-^2 Z_-^2 \sigma_-^2 \hat h_{--}(k) {\rm e}^{-k^2 \sigma_-^2} \right \}
k^2 dk 
\label{eq:p_k}
\end{align}
\end{subequations}
The dimensionless isothermal compressibility $\chi_{\rm T}$ is
determined by the small $k$-limit of the number-number structure
factor:
\begin{equation}
n k_{\rm B} T \chi_{\rm T} = 
\left( \frac{\partial \beta P}{\partial n} \right)^{-1}_{\rm T}
= \lim_{k \to 0} S_{NN}(k) . 
\label{eq:compress}
\end{equation}

The entropy $S$, Helmholtz free energy $F$ or chemical potentials
$\mu_+$ and $\mu_-$ cannot be expressed in terms of the pair
distribution functions alone. The dimensionless excess Helmholtz free
energy per polyion, $f^{\rm ex} = \beta F^{\rm ex}/N$, along an
isochore can be obtained by standard thermodynamic integration,
starting from the high temperature ($\beta = 0$) limit, where $f^{\rm
  ex} = 0$, according to
\begin{equation}
f^{\rm ex}(\beta,  n) = 
\int_0^\beta u^{\rm ex}(\beta ', n) \frac{d \beta '}{\beta '} .
\end{equation}
Equivalently, the chemical potentials of the polyions may be
calculated by the Kirkwood charging process~\cite{kirkwood_1935} whereby a test
polycation or polyanion is gradually coupled to the $N$ polyions of
the system by varying a coupling parameter $\lambda$ from 0 (non
interacting test particle) to 1 (fully interacting test particle):
\begin{equation}
\mu^{\rm ex}_\alpha = n \int_0^1 d \lambda \int \sum_\beta x_\beta v_{\alpha \beta}(r)
h_{\alpha \beta}(r; \lambda) d {\bf r} 
\label{eq:mu_ex}
\end{equation}
where $h_{\alpha \beta}(r; \lambda)$ is the pair correlation function
between the test particle of species $\alpha$ and the bath particles
of species $\beta$ corresponding to a pair potential $\lambda
v_{\alpha \beta}(r)$.


Because of the absence of strong short-range interactions and the
long-range nature of the Coulombic interactions, a ``natural''
approximation for the pair correlation functions is provided by the
random phase approximation (RPA), which is expected to be very
accurate at high temperatures ($T \gtrsim 1$) and high densities. The
RPA amounts to setting the direct correlation functions equal to their
asymptotic limit, for all inter-particle distances $r$, namely
\begin{subequations}
\begin{align}
c_{\alpha \beta}(r) & = - \beta v_{\alpha \beta}(r) = 
- Z_\alpha Z_\beta \frac{\Gamma}{r} {\rm erf}(r/2 \sigma_{\alpha \beta}) \\
\hat c_{\alpha \beta}(k) & = - \beta \hat v_{\alpha \beta}(k) = 
- \frac{4 \pi Z_\alpha Z_\beta \Gamma n}{k^2} {\rm e}^{-k^2 \sigma_{\alpha \beta}^2} 
~~~~~ \alpha = +, -. 
\label{eq:c_rpa_k}
\end{align}
\end{subequations}
Substitution of Eq.~(\ref{eq:c_rpa_k}) into Eqs.~(\ref{eq:def_h_pp})
to (\ref{eq:def_delta_d}) leads to simple analytic expressions for the
$\hat h_{\alpha \beta}(k)$, while Eqs.~(\ref{eq:def_s_nn}) to
(\ref{eq:def_s_cc}) yield the following explicit expressions for the
structure factors
\begin{subequations}
\begin{align}
\nonumber
S_{NN}(k) & = 1 - \frac{4 \pi n l_{\rm B}}{k^2 D(k)}
\left[ x_+^2 Z_+^2 {\rm e}^{-k^2 \sigma_+^2} + x_-^2 Z_-^2 {\rm e}^{-k^2 \sigma_-^2} \right. \\
& \left. + 2 x_+ x_- Z_+ Z_- {\rm e}^{-k^2 \sigma_{+-}^2}  \right] \\
\nonumber
S_{NC}(k) & = - \frac{4 \pi n l_{\rm B}}{k^2 D(k)}
\left[ x_+^2 Z_+^3 {\rm e}^{-k^2 \sigma_+^2} + x_-^2 Z_-^3 {\rm e}^{-k^2 \sigma_-^2} \right. \\
& \left. + x_+ x_- Z_+ Z_- (Z_+ + Z_-){\rm e}^{-k^2 \sigma_{+-}^2}  \right] \\
\nonumber
S_{CC}(k) & = \bar Z^2 - \frac{4 \pi n l_{\rm B}}{k^2 D(k)}
\left[ x_+^2 Z_+^4 {\rm e}^{-k^2 \sigma_+^2} + x_-^2 Z_-^4 {\rm e}^{-k^2 \sigma_-^2} \right. \\
& \left. + 2 x_+ x_- Z_+^2 Z_-^2 {\rm e}^{-k^2 \sigma_{+-}^2}  \right]  
\label{eq:s_cc_rpa}
\end{align}
\end{subequations}
where according to (\ref{eq:def_delta_d}) and (\ref{eq:c_rpa_k})
\begin{equation}
D(k) = 1 + \frac{\kappa_{\rm D}^2}{\bar Z^2 k^2} 
\left[ x_+ Z_+^2 {\rm e}^{-k^2 \sigma_+^2} +
x_- Z_-^2 {\rm e}^{-k^2 \sigma_-^2}  \right] .
\end{equation}
It is easily verified that $\lim_{k \to 0} S_{NN}(k) = 1$, so that,
according to Eq.~(\ref{eq:compress}) the isothermal
compressibility is that of an ideal gas, a well-known deficiency of
the RPA. Similarly, $\lim_{k \to 0} S_{NC}(k) = 0$, so that the RPA
predicts that charge and number density fluctuations are decoupled in
the long wavelength limit. After some algebra the charge-charge
structure factor (\ref{eq:s_cc_rpa}) may be cast in the convenient
form
\begin{equation}
\frac{S_{CC}(k)}{\bar Z^2} =
\frac{
\bar Z^2 k^2 + x_+x_- Z_+^2 Z_-^2 \kappa_D^2 \left[ {\rm e}^{-k^2 \sigma_+^2/2} - 
{\rm e}^{-k^2 \sigma_-^2/2} \right]
}
{
\bar Z^2 k^2 + \kappa_D^2 \left[ x_+ Z_+^2 {\rm e}^{-k^2 \sigma_+^2} + 
x_- Z_-^2 {\rm e}^{-k^2 \sigma_-^2} \right]
} 
\label{eq:scc_rpa_2}
\end{equation}
which clearly satisfies the Stillinger-Lovett condition
(\ref{eq:stillinger_lovett_limit}).

In the remainder of this paper we focus on the symmetric (restricted)
version of the UPM, namely the URPM ($Z_+ = - Z_- = Z$; $n_+ = n_- =
n/2$; $\sigma_+ = \sigma_- = \sigma_{+-} \equiv \sigma$). In this
case, $v_{++}(r) \equiv v_{--}(r) = - v_{+-}(r)$, and hence $g_{++}(r)
\equiv g_{--}(r)$, so that there are only two independent pair
distribution functions and structure factors. Furthermore, all the
relations above simplify considerably for the URPM. For instance,
$S_{NN}(k) \equiv 1$, $S_{NC}(k) \equiv 0$ and $S_{CC}(k)$ reduces to
\begin{equation}
\frac{S_{CC}(k)}{\bar Z^2} = \frac{k^2}{k^2 + \kappa_D^2 {\rm e}^{-k^2
    \sigma^2}}
\label{eq:scc_rpa_urpm}
\end{equation}
For point ions, $S_{CC}(k)$ goes over to the standard Debye-H\"uckel
form. Note that, within RPA, $h_{++}(r) = - h_{+-}(r) =
h_{CC}(r)/2$. RPA estimates of the thermodynamic properties are
obtained by substituting the above expressions for the pair
correlation functions in Eqs.~(\ref{eq:u_ex_k}) and
(\ref{eq:p_k}). The exact Kirkwood expression (\ref{eq:mu_ex}) for the
excess chemical potentials $\mu_+ = \mu_- \equiv \mu$ reduces to
\begin{eqnarray} \nonumber
\beta \mu^{\rm ex} & = & 
\frac{n}{2} \int_0^1 d \lambda \int \beta v(r) h_{CC}(r; \lambda) d {\bf r} \\ 
& = & \frac{1}{2} \frac{1}{(2 \pi)^3} 
\int_0^1 d \lambda \int \beta \hat v(k) \hat h_{CC}(k; \lambda) d {\bf k}
\label{eq:mu_urpm}
\end{eqnarray}
where $v(r) \equiv v_{++}(r)$, $\beta \hat v(k) = \frac{4 \pi Z^2
  \Gamma}{k^2} {\rm e}^{-k^2 \sigma^2}$, and $h_{CC}(r) = h_{++}(r) -
h_{+-}(r)$. 

Using the coupled OZ relations for the correlation functions of a test
particle partially coupled to the bath particles one easily shows that
\begin{equation}
\hat h_{CC}(k; \lambda) = \frac{\hat c_{CC}(k; \lambda)}{1 - \frac{1}{2} \hat c_{CC}(k)}
\end{equation}
where $\hat c_{CC}(k) = \hat c_{++}(k) - \hat c_{+-}(k) = - 2 n \beta
\hat v(k)$ and $\hat c_{CC}(k; \lambda) = - 2 \lambda \beta \hat v(k)
\equiv \lambda \hat c_{CC}(k)$ within RPA. Hence $\hat h_{CC}(k;
\lambda) = \lambda \hat h_{CC}(k)$ so that the integration over
$\lambda$ in Eq.~(\ref{eq:mu_urpm}) is trivial leading to
\begin{eqnarray}
\nonumber
\beta \mu^{\rm ex} & = & \frac{1}{4} \frac{1}{(2 \pi)^3}
\int \hat h_{CC}(k) \beta \hat v(k) d {\bf k} \\
& = & \frac{\Gamma Z^2}{2 \pi} \int_0^\infty \hat h_{CC}(k) {\rm e}^{-k^2 \sigma^2} dk
\equiv u^{\rm ex} .
\label{eq:mu_urpm_rpa}
\end{eqnarray}
The last equality follows from Eq.~(\ref{eq:u_ex_k}) adapted to the
symmetric case. The result that the excess chemical potential is equal
to the excess internal energy per ion is of course only true within
RPA.

It is easily verified from (\ref{eq:mu_urpm_rpa}) and
(\ref{eq:scc_rpa_urpm}) that the low temperature limit of the RPA
internal energy per ion coincides with the ground state energy
(\ref{eq:u_0})
\begin{equation}
\lim_{T \to 0} \frac{U^{\rm RPA}(T)}{N} = - \frac{N}{2} u = \frac{U_0}{N} .
\end{equation}
In the high temperature limit the leading contribution to the reduced
excess energy per ion reduces to the Debye-H\"uckel limiting law for
point ions
\begin{equation}
\lim_{T \to \infty} \frac{\beta U^{\rm RPA}}{N} = - \frac{\kappa_D^3}{8 \pi n} .
\end{equation}
In the zero temperature limit the pressure is negative
\begin{equation}
\lim_{T \to 0} \frac{P^{\rm RPA}}{n} = - \frac{u}{12} .
\end{equation}

Among the standard integral equations for the pair structure the
hypernetted chain (HNC) equation is known to be well adapted to
Coulombic fluids~\cite{hansen_theory_2006}. HNC theory supplements the coupled OZ relations
linking the total and direct correlation functions $h_{\alpha
  \beta}(r)$ and $c_{\alpha \beta}(r)$, by the closure relations
\begin{eqnarray}
\nonumber
h_{\alpha \beta}(r)+1 = g_{\alpha \beta}(r) & =  &
  \exp \left \{ - \beta v_{\alpha \beta}(r) + h_{\alpha \beta}(r) - c_{\alpha \beta}(r) \right\} \\
& = & \exp \left \{ h_{\alpha \beta}(r) - \Delta c_{\alpha \beta}(r) \right \} 
\label{eq:hnc}
\end{eqnarray}
where $\Delta c_{\alpha \beta}(r) = c_{\alpha \beta}(r) + \beta
v_{\alpha \beta}(r)$ are the short-range parts of the direct
correlation functions, which are expected to vanish rapidly for large
$r$ (within the RPA, $\Delta c_{\alpha \beta}(r) \equiv 0$). In the
symmetric case under consideration here, $h_{++}(r) \equiv h_{--}(r)$,
$c_{++}(r) \equiv c_{--}(r)$, and the three coupled OZ relations
reduce to two decoupled equations for the number-number
and charge-charge correlation functions $h_{NN}(r) = h_{++}(r) +
h_{+-}(r)$, $h_{CC}(r) = h_{++}(r) - h_{+-}(r)$; in $k$-space the two
relations read
\begin{equation}
\hat h_{\alpha \alpha}(k) = 
\frac{\hat c_{\alpha \alpha}(k)}{1 - \frac{1}{2} \hat c_{\alpha \alpha}(k)}; 
\quad \alpha = N, C .
\label{eq:h_hnc}
\end{equation}
HNC theory has the advantage that the excess chemical potential can be
calculated from a knowledge of the pair correlation functions
alone~\cite{hansen_theory_2006}. The set of Eq.~(\ref{eq:hnc}) and (\ref{eq:h_hnc}) are
solved numerically by an iterative Picard method, taking particular
care of the long-range tails of the direct correlation functions in
$r$- and $k$-space. Unfortunately, the range of thermodynamic
conditions for which the numerical procedure converges is limited to
increasingly higher temperatures as the density is reduced. The
applicability of the HNC closure is therefore limited to relatively
high density and temperatures (see below).

\begin{figure}
\begin{center}
\includegraphics[width=\onefig,clip=true,draft=false,angle=0]
               {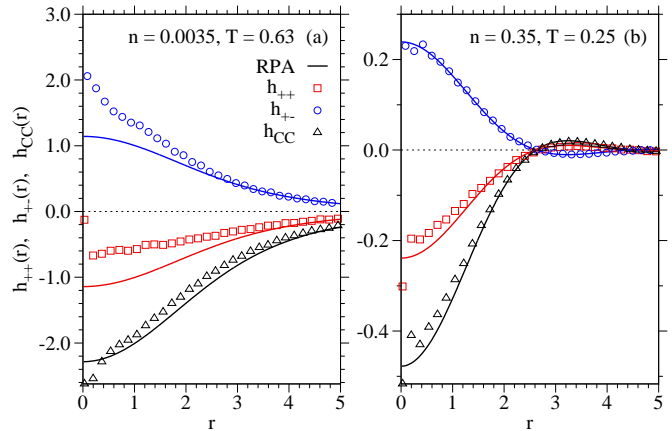}
\end{center}
\caption{Total correlation functions $h_{++}(r)$, $h_{+-}(r)$ and
  $h_{CC}(r)$ from MC simulations (symbols) and RPA (full lines) for
  two different state points: (a) $n=0.0035$, $T=0.63$ and (b)
  $n=0.35$, $T=0.25$. }
\label{fig:1}
\end{figure}

\begin{figure}
\begin{center}
\includegraphics[width=\onefig,clip=true,draft=false,angle=0]
                {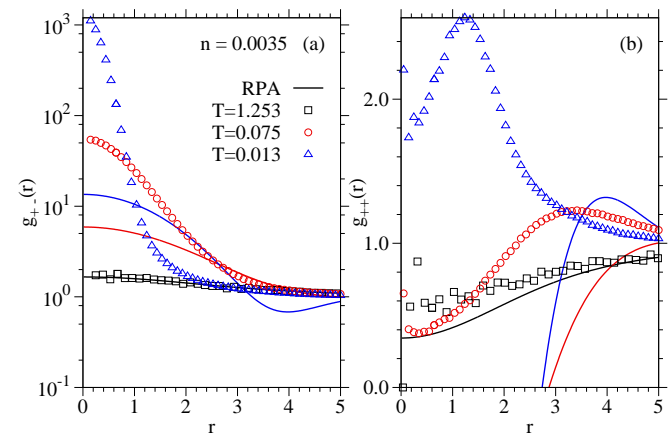}
\end{center}
\caption{Pair distribution functions along the isochore $n = 0.0035$
  for three different temperatures (as specified) from MC simulations
  (symbols) and RPA (full lines): (a) $g_{+-}(r)$ and (b)
  $g_{++}(r)$.}
\label{fig:PDFs_0.01}
\end{figure}

\begin{figure}
\begin{center}
\includegraphics[width=\onefig,clip=true,draft=false,angle=0]
                {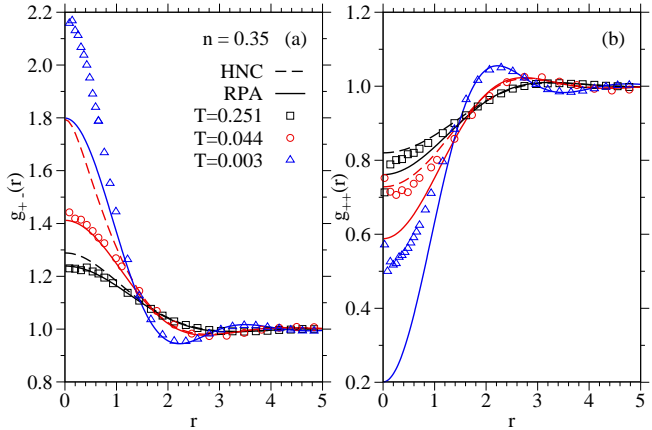}
\end{center}
\caption{Pair distribution functions along the isochore $n = 0.35$
  for three different temperatures (as specified) from MC simulations
  (symbols), RPA (full lines), and HNC (dashed lines): (a) $g_{+-}(r)$
  and (b) $g_{++}(r)$.}
\label{fig:PDFs_1.00}
\end{figure}

\begin{figure}
\begin{center}
\includegraphics[width=\onefig,clip=true,draft=false,angle=0]
                {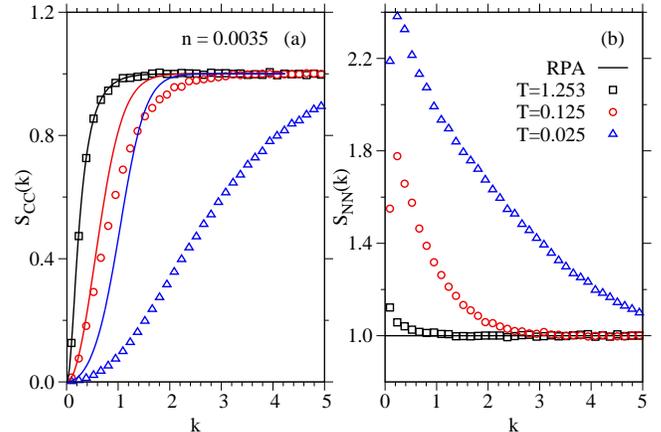}
\end{center}
\caption{Structure factors along the isochore $n=0.0035$ for three
  different temperatures (as specified) from MC simulations (symbols)
  and RPA (full lines): (a) $S_{CC}(k)$ and (b) $S_{NN}(k)$.}
\label{fig:SKs_0.01}
\end{figure}

We have compared the RPA and HNC predictions for the pair structure of
the URPM to MC results along several isochores. The details of our simulations are described in Appendix~\ref{sec:appendix_a}. In the following we
focus on the two isochores $n=0.0035$ and $n=0.35$, which are
representative of the system's behavior at low and high density,
respectively. Note that we do not investigate here the regime of very
low densities, in which special simulation techniques must be employed
to ensure ergodicity~\cite{valeriani_computer_2010}. Data along the
isochores $n=0.0035$ and $n=0.35$ are shown in Fig.~\ref{fig:1} to
\ref{fig:SKs_0.01}, respectively. Figure~\ref{fig:1} shows
$h_{++}(r)$, $h_{+-}(r)$ and $h_{CC}(r)$ for $n=0.0035$, $T = 0.63$ and
$n=0.35$, $T = 0.25$. Since these temperatures are below the no-solution line of HNC, we only report the predictions of RPA. At the
higher density, RPA is seen to be very accurate for $h_{CC}(r)$,
although the RPA symmetry $h_{+-}^{\rm RPA}(r) = - h_{++}^{\rm
  RPA}(r)$ is broken in the simulation data. The discrepancies between
RPA and simulation data are only slightly more pronounced at the lower
density ($n=0.0035$, $T = 0.63$), but they increase rapidly as $T$ is
lowered due to strong Coulomb correlations (see below).

A comparison between theoretical predictions and MC data at even lower
temperatures is made in Fig.~\ref{fig:PDFs_0.01} along the isochore $n
= 0.0035$, and along the isochore $n=0.35$ in
Fig.~\ref{fig:PDFs_1.00}. Strong pairing is evident from inspection of
the data for $g_{+-}(r)$ at the lower densities and temperatures as
expected. Indirect evidence of pairing is also provided by the
observation that at the lowest temperatures, along the isochore $n =
0.0035$, $g_{++}(r) > 1$ as $r \to 0$ despite the Coulomb repulsion
between equal sign polyions; this apparent attraction can be
rationalized by the formation of tight anion/cation pairs: the anion
of a given pair will be preferentially situated between its cationic
partner and the cation of another pair, thus favoring the
cation-cation approach and the possible formation of trimers and
higher order clusters. RPA is unable to properly account for strong
ion pairing, both at high and low density. The agreement of HNC
predictions with the MC data at high density (see
Fig.~\ref{fig:PDFs_1.00}) is good at high temperature but is seen to
deteriorate as the temperature is lowered towards the
no-solution point (corresponding to $T=0.044$ at $n=0.35$). Also note the
unphysical predictions of RPA ($g_{++}(r)<0$) at low density and
temperature.

\begin{figure}
\begin{center}
\includegraphics[width=\onefig,clip=true,draft=false,angle=0]
                {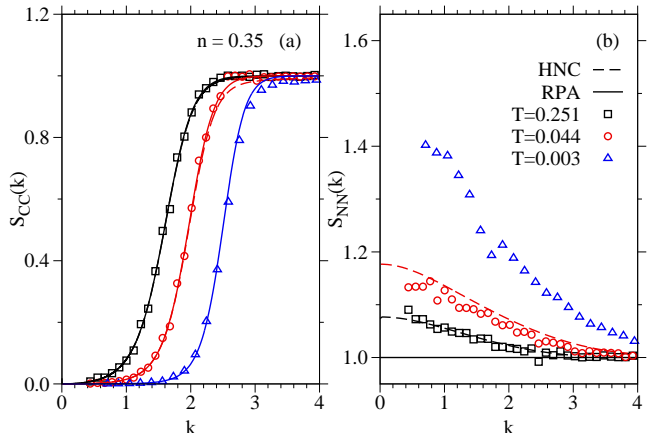}
\end{center}
\caption{Structure factors along the isochore $n=0.35$ for three
  different temperatures (as specified) from MC simulations (symbols),
  RPA (full lines) and HNC (dashed lines): (a) $S_{CC}(k)$ and (b)
  $S_{NN}(k)$.}
\label{fig:SKs_1.00}
\end{figure}

Figure~\ref{fig:SKs_0.01}-a shows a comparison between the RPA
predictions and MC data for the charge-charge structure factor
$S_{CC}(k)$ along the low density isochore $n=0.0035$. While the
agreement is very good at the higher temperatures ($T \gtrsim 0.2$),
it deteriorates rapidly as $T$ is lowered. At the lowest temperatures
($T \lesssim 0.05$) the MC data do not satisfy the Stillinger-Lovett
condition (\ref{eq:stillinger_lovett_limit}), while the RPA result
(\ref{eq:scc_rpa_2}) obeys the condition by construction. As shown
in~\cite{coslovich_clustering_2011}, a least squares parabolic fit to
the low-$k$ MC data [$S_{CC}(k)/Z ^2 \simeq k^2/\kappa^2$] yields an
effective inverse Debye screening length $\kappa$ which is
systematically larger than $\kappa_{\rm D}$ [defined in
Eq.~(\ref{eq:kappa_d})]. The violation of the Stillinger-Lovett limit
suggests that the URPM no longer behaves as a purely ionic fluid at
low temperatures, due to ion pairing; we will return to this important
issue in Sec.~\ref{sec:clustering_dielectric}, where clustering and
dielectric response will be analyzed. A similar analysis of
$S_{CC}(k)$ along the high density isochore (see
Fig.~\ref{fig:SKs_1.00}) shows excellent agreement between RPA and MC
data down to $T = 0.003$, with the effective $\kappa$ consistently
close to $\kappa_{\rm D}$, confirming that the URPM remains ionic
throughout: at high densities anion/cation pairs are short-lived due
to the frequent overlap of neighboring pairs.

The results for $S_{NN}(k)$ are shown in Fig.~\ref{fig:SKs_0.01}-b and
Fig.~\ref{fig:SKs_1.00}-b. The density-density structure factor
$S_{NN}(k)$ is seen to increase rapidly as $k \to 0$ along the
isochore $n=0.0035$, signaling the proximity of a spinodal line as
$T$ is lowered. At the spinodal line associated with a phase
separation, $S_{NN}(k)$ diverges in the long wavelength limit. Note
that, within RPA, $S_{NN}(k)\equiv 1$. The predictions of HNC along the
high density isochore indicate a mild increase at small $k$,
consistent with the simulation data. However, the integral equation approach
admits no solution in the vicinity of phase coexistence---a well-known
defect of the theory~\cite{fisher_1981,belloni_1993}. All these findings clearly
indicate that more sophisticated theoretical tools should be employed to study the
low temperature, low density regime of the model.

\begin{figure}
\begin{center}
\includegraphics[width=\onefig,clip=true,draft=false,angle=0]
                {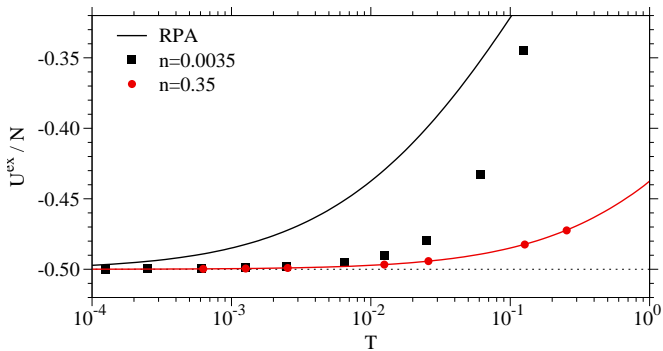}
\end{center}
\caption{Excess internal energy per particle $U^{\rm ex}/N$ as a
  function of temperature from MC simulations (symbols) and RPA (full lines) along the isochores $n=0.0035$ (squares) and
  $n=0.35$ (circles).}
\label{fig:U}
\end{figure}

\begin{figure}
\begin{center}
\includegraphics[width=\onefig,clip=true,draft=false,angle=0]
                {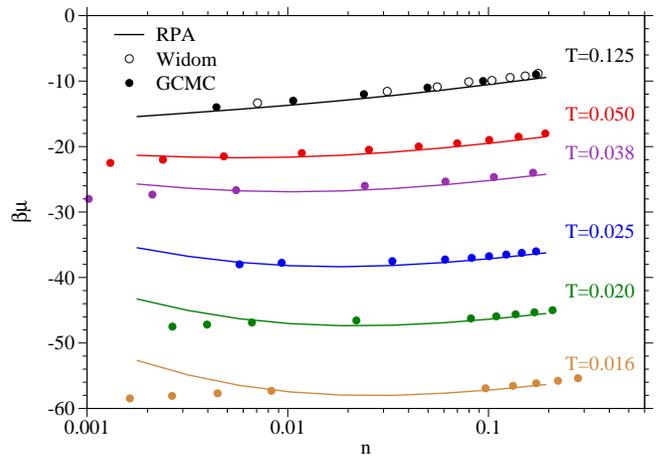}
\end{center}
\caption{Dimensionless chemical potential $\beta \mu$ as a function
  of the density $n$ along several isotherms (as labeled) obtained from the
  Widom insertion method in NVT-MC simulations (open circles), grand
  canonical MC simulations (full circles) and RPA (full lines).}
\label{fig:mu}
\end{figure}

The RPA, HNC and MC data for the pair structure may be used to compute
the excess internal energy and the equation of state of the URPM via
Eqs.~(\ref{eq:u_ex_r}), (\ref{eq:u_ex_k}) and (\ref{eq:p_r}),
(\ref{eq:p_k}), and the isothermal compressibility via
Eq.~(\ref{eq:compress}). The excess chemical potential follows from
Eq.~(\ref{eq:mu_urpm}), leading to the explicit result
(\ref{eq:mu_urpm_rpa}) within the RPA. 
Representative comparisons are made for $u^{\rm ex}(n, T) = U^{\rm
  ex}/N$ in Fig.~\ref{fig:U} along the
isochores $n=0.0035$ and $n = 0.35$, and for the total, dimensionless
chemical potential $\beta \mu$ in Fig.~\ref{fig:mu}. Simulation results for the density
dependence of $\beta \mu$ have been obtained from
grand-canonical MC simulations using biased pair
insertions~\cite{orkoulas_free_1994}. Also included are results for
$\beta \mu$ obtained from standard NVT MC simulations using
the Widom insertion method~\cite{widom_1963}. The agreement between
RPA and simulation results is good only at sufficiently high
temperature and density. We note that for these temperatures and
densities, RPA- and HNC-data agree within symbol size. For
completeness, we note that the Helmholtz excess free energy per ion
can finally be obtained from the thermodynamic relation
\begin{equation}
f^{\rm ex} = \frac{\beta F^{\rm ex}}{N} = \beta \mu^{\rm ex} -
\frac{\beta P^{\rm ex}}{n} \,.
\end{equation}

\section{Clustering and dielectric response}
\label{sec:clustering_dielectric}

The structural data and thermodynamic properties reported in section
\ref{sec:structure_thermo} point to strong clustering of anions and
cations at low temperatures and densities. Pairing of oppositely
charged ions, as well as the formation of larger aggregates are well
documented for the RPM
\cite{bjerrum__1926,stillinger_ion-pair_1968,gillan_liquid-vapour_1983,levin_criticality_1996,valeriani_ion_2010}, where
clusters carry large multipole moments (dipoles in the case of pairs)
due to the ionic cores which induce charge separation. In the case of
the URPM opposite charges can overlap, so that clusters do not carry
permanent multipoles, but are polarizable entities. Their mutual
interactions are thus expected to be much weaker than in the case of
the RPM.

\begin{figure}
\begin{center}
\includegraphics[width=\onefig,clip=true,draft=false,angle=0]
                {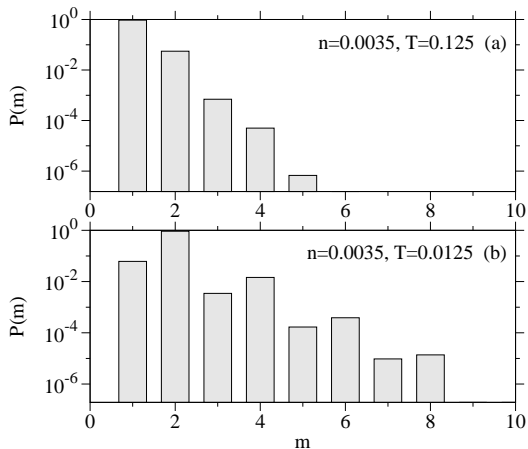}
\end{center}
\caption{Cluster distribution function $P(m)$ for (a) $T=0.125$ and
  (b) $T=0.0125$ along the isochore $n=0.0035$.}
\label{fig:distribution}
\end{figure}

Clusters can be properly defined only at sufficiently low densities,
and even then there is always some arbitrariness in their
definition. We have used a standard geometric definition of
$m$-clusters, namely that $m$ ions form an $m$-mer if each ion lies
within a given distance $r_c$ of at least one other ion in the
cluster. The cut-off $r_c$ is taken to be typically close to 1.0 (in
units of $\sigma$), corresponding to a situation where the charge
distributions of polyions touch (cf. Eq.~(\ref{eq:charge_density})),
and we have examined the sensitivity of the cluster analysis to
variations of $r_c$. We have thus determined the cluster distribution
functions, i.e., the fractions $P(m)$ of $m$-mers, averaged over all
configurations generated in MC and MD simulations. Typical examples of
the resulting histograms are shown in Fig.~\ref{fig:distribution} for
$n=0.0035$ and two temperatures. At the higher temperature ($T = 0.125$)
monomers (i.e., free ions) are by far the majority species, and
$P(m)$ is seen to decrease monotonically and rapidly as $m$
increases. At the lower temperature ($T=0.0035$), pairs constitute the
most common $m$-mers (with a fraction $P(2)$ very close to 1), and the
variation with $m$ is alternating, with neutral $m$-mers (pairs,
tetramers, etc.) being much more probable than charged $m$-mers
(corresponding to odd values of $m$).

\begin{figure}
\begin{center}
\includegraphics[width=\onefig,clip=true,draft=false,angle=0]
                {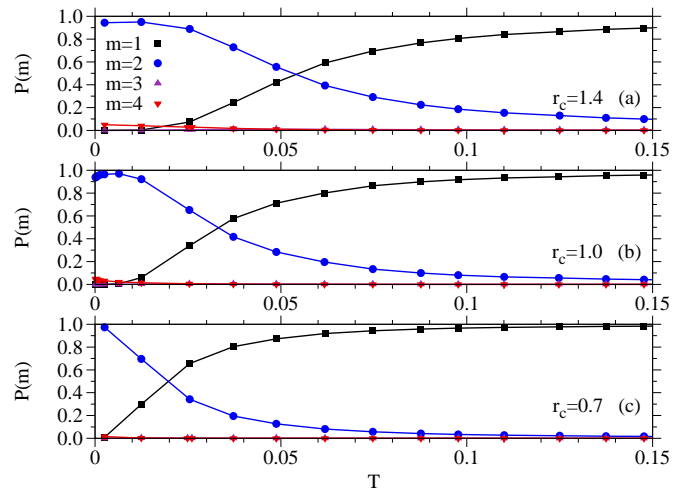}
\end{center}
\caption{Fractions of selected cluster types $P(m)$ as functions of
  temperature along the isochore $n=0.0035$:
  $P(m=1)$ (squares), $P(m=2)$ (circles), $P(m=3)$ (triangles), and
  $P(m=4)$ (reversed triangles). The three panels show results
  obtained using the following values of the cut-off $r_c$: (a)
  $r_c=1.4$, (b) $r_c=1.0$, and (c) $r_c=0.7$.}
\label{fig:cluster_cutoff}
\end{figure}

The variation of $P(m)$ as a function of temperature, for $1 \le m \le
4$, is illustrated in Fig.~\ref{fig:cluster_cutoff} for three
different choices of the cut-off $r_c$. Although there are significant
quantitative differences between the three sets of data, the trends
are the same. As expected the fraction $P(1)$ of monomers is
vanishingly small at the lowest temperatures; it increases rapidly
towards 1 for $T \gtrsim 0.05$. Simultaneously the fraction $P(2)$ of
dimers drops dramatically as $T$ increases; the fraction $P(3)$ of
trimers is negligibly small throughout, while the fraction of
tetramers is non negligible at the lowest temperatures. These trends
agree with intuitive expectation, but our analysis provides a
quantitative estimate of the temperature range associated with the
onset of pairing. The onset of pairing along the isochore $n=0.0035$
is seen to be rapid, but continuous. A similar cluster analysis
carried out along several low density isochores shows that the
temperature at which $P (m=2) \simeq P(m=1) \simeq 0.5$ drops as the density $n$
increases.

\begin{figure}
\begin{center}
\includegraphics[width=\onefig,clip=true,draft=false,angle=0]
                {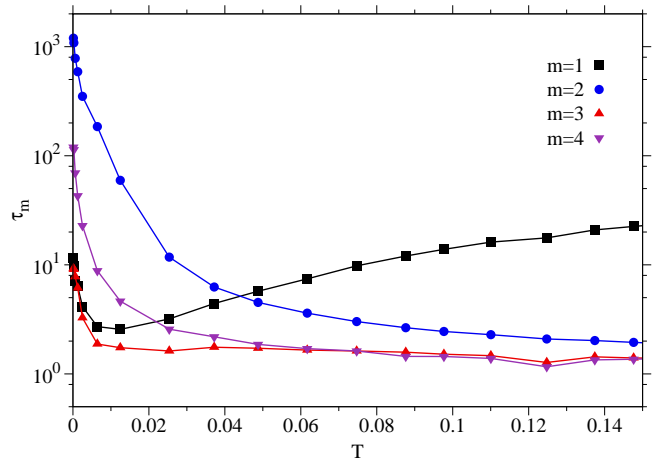}
\end{center}
\caption{Cluster lifetime $\tau_m$ at $n = 0.0035$ as a function of temperature for
  selected cluster types: $m=1$ (squares), $m=2$ (circles),
  $m=3$ (triangles), and $m=4$ (reversed triangles).}
\label{fig:cluster_lifetimes}
\end{figure}

Using MD simulations, we have estimated the mean cluster lifetimes
$\tau_m$ for $m$-mers with $1 \le m \le 4$ as functions of temperature, along the
isochore $n=0.0035$. The lifetime is defined as the minimum time span
during which a cluster is formed by exactly the same set of particles,
and is thus bounded from below by the typical collision time between
clusters.  Results are shown in Fig.~\ref{fig:cluster_lifetimes}. The
monomer lifetime is seen to increase from $\tau_1 \simeq 2$ up to
$\tau_1 \simeq 15$ as the temperature rises, while the opposite trend
of the dimer lifetime is seen to be much faster, with $\tau_2 = 10^3$
at the lowest $T$, dropping rapidly to $\tau_2 \simeq 5$ at $T \simeq
0.05$, the temperature at which $\tau_1$ and $\tau_2$ cross. The
lifetimes $\tau_3$ and $\tau_4$ of trimers and tetramers are even
smaller at high $T$, although they tend to increase in the
low-temperature, paired regime. 

\begin{figure}
\begin{center}
\includegraphics[width=\onefig,clip=true,draft=false,angle=0]
                {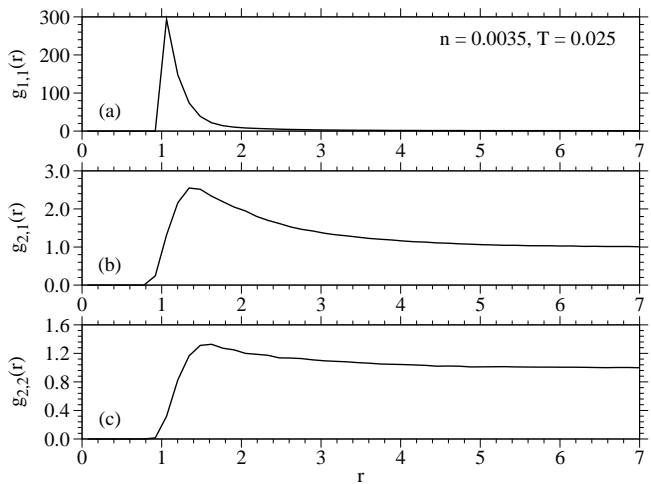}
\end{center}
\caption{Distribution functions $g_{m,n}(r)$ between the centers of
  mass of selected cluster types for $n=0.0035$, $T=0.025$: (a)
  monomer-monomer $g_{1,1}(r)$, (b) monomer-dimer $g_{1,2}(r)$, and (c)
  dimer-dimer $g_{2,2}(r)$.}
\label{fig:PDF_pp_pi}
\end{figure}

At the lower temperatures, where the lifetimes $\tau_2$ of pairs are
long, it makes sense to consider the systems as made up of three
``species'', namely free polyanions, polycations, and neutral
polyanion-polycation pairs (``chemical picture''). Under those
conditions, one may determine monomer-dimer and dimer-dimer
distribution functions $g_{1,2}(r)$ and $g_{2,2}(r)$ between isolated
monomers and dimers that are not part of larger clusters, in MC or MD
simulations. Examples are compared to the monomer-monomer pair
distribution function $g_{1,1}(r)$ in Fig.~\ref{fig:PDF_pp_pi}, for
$n=0.0035$ and $T = 0.025$, i.e., close to the pairing
transition. $g_{1,1}(r)$ is dominated by the strong anion-cation
attraction, leading to a pronounced peak at $r = 1$. $g_{1,2}(r)$ and
$g_{2,2}(r)$ are seen to exhibit modest peaks at slightly larger
distances pointing to rather weak correlations. Note that all three
pair distribution functions vanish for $r \alt r_c=1.0$ by construction,
because for shorter CM-CM distances the two monomers would be
identified as a single dimer, a monomer and a dimers would be
identified as a single trimer, and two dimers would be identified as a
single tetramer, implying an apparent infinite barrier~\cite{stillinger_ion-pair_1968}.

\begin{figure}
\begin{center}
\includegraphics[width=\onefig,clip=true,draft=false,angle=0]
                {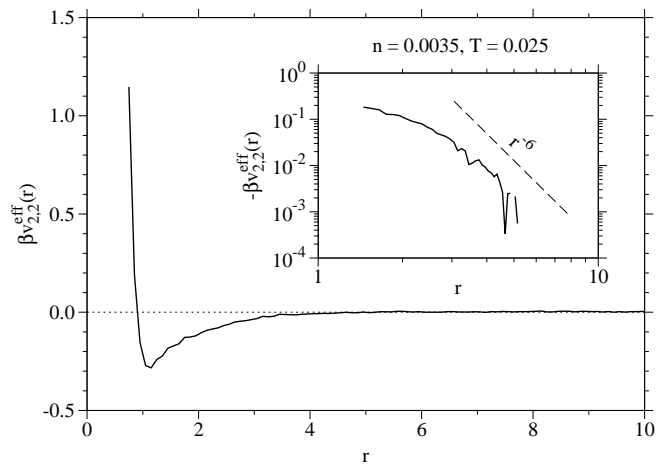}
\end{center}
\caption{Dimensionless, effective dimer-dimer potential $\beta
  v_{2,2}(r)$ for $n=0.0035$, $T=0.025$. Inset: effective potentials
  on a double-logarithmic scale. The dashed line indicates the
  asymptotic behavior expected at large $r$ (see
  Appendix~\ref{sec:appendix_b}).}
\label{fig:effective_potentials_pp}
\end{figure}

\begin{figure}
\begin{center}
\includegraphics[width=\onefig,clip=true,draft=false,angle=0]
                {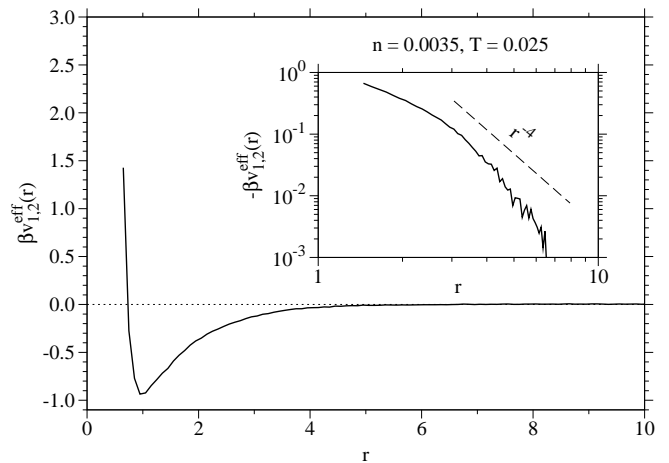}
\end{center}
\caption{Same as Fig.~\ref{fig:effective_potentials_pp} but for the
  monomer-dimer effective potential $\beta v_{1,2}(r)$.}
\label{fig:effective_potentials_pi}
\end{figure}

In view of the low density, the pair distribution function data may be
inverted to determine effective monomer-monomer and monomer-dimer pair
potentials $v_{1,2}(r)$ and $v_{2,2}(r)$ by a simple Boltzmann
inversion
\begin{equation}
v_{n, m}(r) = - k_{\rm B} T \ln g_{n, m}(r) .
\end{equation}
The resulting effective potentials are pictured in
Figs.~\ref{fig:effective_potentials_pp}
and~\ref{fig:effective_potentials_pi}. As explained above, the
short-range repulsion is an artifact linked to cluster
identity. Beyond $r = r_c$ the monomer-dimer and dimer-dimer
potentials are seen to be weak and attractive. In view of the statistical uncertainties, the asymptotic
behaviors at large distances, illustrated by the insets in
Figs.~\ref{fig:effective_potentials_pp} and~\ref{fig:effective_potentials_pi},
are in reasonable agreement with the theoretical predictions $v_{1,2}(r)
\sim 1/r^4$ and $v_{2,2}(r) \sim 1/r^6$ for isolated monomer-dimer and
dimer-dimer pairs (see Appendix B). These are of course the same
asymptotic behaviors as for van der Waals-London dispersion
interactions between polarizable atoms or molecules, but the
prefactors are proportional to $k_{\rm B} T$ in the present, purely
classical case, while they are independent of temperature in the
atomic case, since they result from quantum averages over the
electronic ground state.

Another diagnostic for clustering is provided by the dielectric
response of the URPM. We have shown in Section
\ref{sec:structure_thermo} that at high temperatures and densities the
system behaves as a conductor, obeying the Stillinger-Lovett perfect
screening condition Eq.~\eqref{eq:stillinger_lovett_limit}. However, as
illustrated in Fig.~\ref{fig:SKs_0.01}, the Stillinger-Lovett
condition is violated at low densities and temperatures pointing to a
dielectric, rather than conducting behavior. According to static
linear response theory~\cite{hansen_theory_2006}, the charge-charge
structure factor is related to the $k$-dependent dielectric response
function by
\begin{equation}\label{eqn:kappa}
\frac{1}{\epsilon(k)} = 
1 - \frac{\kappa_{\rm D}^2}{k^2} \frac{S_{CC}(k)}{\bar Z^2} .
\end{equation}
Taking the $k \to 0$ limit, where $S_{CC}(k) \simeq \bar Z^2
k^2/\kappa^2$ we arrive at the following expression for the
macroscopic dielectric permittivity $\epsilon$
\begin{equation}
\frac{1}{\epsilon} = \lim_{k \to 0} \frac{1}{\epsilon(k)} = 
1 - \frac{\kappa_{\rm D}^2}{\kappa^2} .
\end{equation}
In a conducting medium $\kappa \equiv \kappa_{\rm D}$, and $\epsilon
\to \infty$ as expected. At sufficiently low $T$ and $n$, all polyions
are paired and the MC data show that $\kappa > \kappa_{\rm D}$, so
that $\epsilon$ takes on a finite value characteristic of an
insulating, dielectric medium
\begin{equation}
\infty > \epsilon = \frac{1}{1 - \kappa_{\rm D}^2/\kappa^2} > 1 .
\end{equation}

Equivalently, $\epsilon$ may be estimated form Kirkwood's fluctuation
formula~\cite{kirkwood_1939} adapted to simulations carried out under
metallic boundary conditions~\cite{neumann_1983}
\begin{equation}
\epsilon = 1 + \frac{4 \pi}{3 k_{\rm B} T} 
\frac{\langle | {\bf M} |^2 \rangle - |\langle  {\bf M} \rangle|^2}{V}
\label{eq:permittivity}
\end{equation}
where ${\bf M} = \sum_i Q_i {\bf r}_i$ is the total dipole moment of
the system.

\begin{figure}
\begin{center}
\includegraphics[width=\onefig,clip=true,draft=false,angle=0]
                {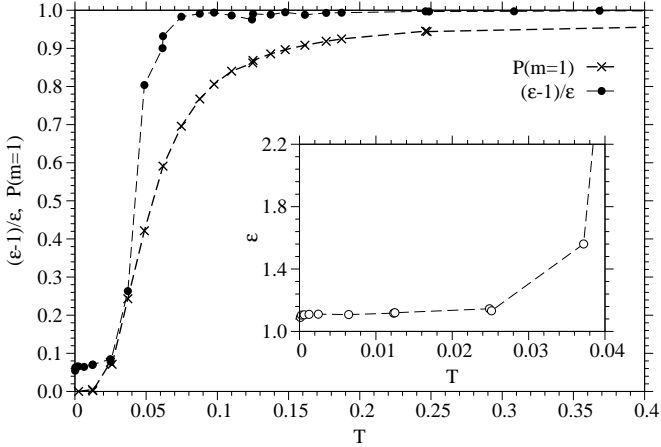}
\end{center}
\caption{Dielectric order parameter $(\epsilon-1)/\epsilon$ (filled
  circles) obtained from Eq.~\eqref{eq:permittivity} and fraction of
  free ions $P(m=1)$ (crosses) at $n=0.0035$ as functions of
  temperature. Inset: enlarged view of the low-temperature behavior of
  $\epsilon$.}
\label{fig:permittivity}
\end{figure}

From the permittivities estimated using Eq.~(\ref{eq:permittivity}) it
is possible to calculate a ``dielectric order parameter'', defined as
$(\epsilon-1)/\epsilon$, which equals $1$ in the conducting phase and
is close to 0 in the insulating phase. The values of
$(\epsilon-1)/\epsilon$ obtained from MC simulations along the
isochore $n=0.0035$ are shown in Fig.~\ref{fig:permittivity} together
with the corresponding percentage of monomers $P(m=1)$. The dielectric
order parameter signals a sharp transition around $T\approx 0.04$,
which correlates with the rapid increase of the fraction of free ions. In the inset
of Fig.~\ref{fig:permittivity} we see that $\epsilon$ remains nearly
constant up to $T \simeq 0.025$, beyond which $\epsilon$ increases
sharply towards a high temperature limit close to $\epsilon \simeq
10^3$ (cf. Ref.~\cite{coslovich_clustering_2011}), typical of a finite
conducting medium (for an infinite conductor, $\epsilon \to \infty$).
The correlation between $(\epsilon-1)/\epsilon$ and $P(m=1)$ is striking
and validates the picture of a transition from a low temperature
dielectric (insulator) state to a high temperature ionic (conductor)
state driven by the break-up of ion pairs. The ``conductor-insulator''
(CI) transition is seen to be fast, but continuous, which may well be
a finite size effect. Simulations on larger systems and a finite size
scaling analysis will be required to find out if the transition
becomes discontinuous in the thermodynamic limit.

The above analysis has been repeated along several low density
isochores, and the locus of points in the $(n, T)$ plane where $P(m=1) =
0.5$ yields an estimate of pairing (or CI) transition line which is
shown in Fig.~\ref{fig:phase_diagram}. In the portion of the phase diagram studied herein, the transition is seen to take
place at lower temperatures as the density increases, contrary to what
could be expected from a simple ``chemical equilibrium''
picture~\cite{levin_criticality_1996} and to what is found in the
RPM~\cite{valeriani_ion_2010}.

\section{Polyion dynamics}
\label{sec:dynamics}

Time-dependent correlation functions are readily calculated by MD
simulations, and provide a quantitative characterization of
single-particle and collective ion dynamics as well as giving access
to linear transport coefficients, like ion mobility and electrical
conductivity. The latter provides an unambiguous diagnostic of a CI
transition. We note that Brownian Dynamics simulations would be more
appropriate for the system at hand, since they account for solvent
effects. However, since we are interested in qualitative, rather than
quantitative aspects of the dynamics, we prefer to employ here the
more efficient MD simulations, which are capable of exploring longer
time windows.

\begin{figure}
\begin{center}
\includegraphics[width=\onefig,clip=true,draft=false,angle=0]
                {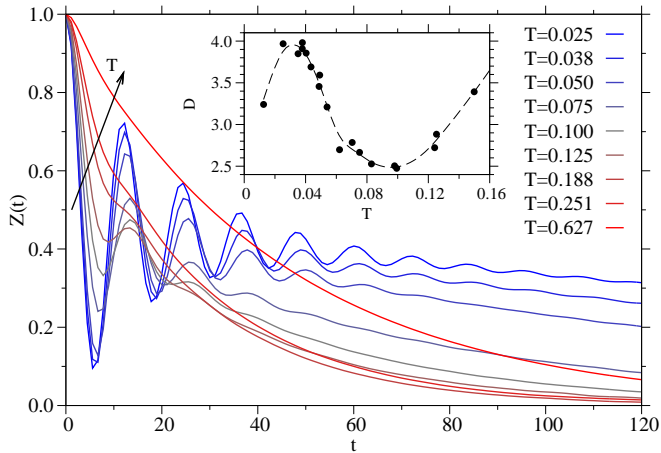}
\end{center}
\caption{Velocity auto-correlation function $Z(t)$, as defined in
  Eq.~(\ref{eq:vacf}), at $n=0.0035$ for different temperatures (as labeled). Inset:
  self diffusion constant $D$ evaluated via
  Eq.~(\ref{eq:self_diffusion}) as a function of temperature.}
\label{fig:vacf}
\end{figure}

Let ${\bf r}(t)$ and ${\bf v}(t)$ denote the position and the velocity
of any given polyion at time $t$; the normalized velocity
auto-correlation functions of the anions and cations are identical for
the symmetric URPM and defined by
\begin{equation}
Z_+(t) = Z_-(t) = Z(t) = \frac{\langle {\bf v}(t) \cdot {\bf v}(0) \rangle}
{\langle v^2 \rangle}
\label{eq:vacf}
\end{equation}
where ${\bf v}(0)$ denotes the initial velocity, $\langle v^2 \rangle
= 3 k_{\rm B} T/m$, and statistical averages are taken along the
trajectories of individual ions, and over all $N$ ions. Examples of
MD-generated correlation functions $Z(t)$, for $n=0.0035$ and four
temperatures, are shown in Fig.~\ref{fig:vacf}. At the higher
temperatures, $Z(t)$ is seen to decay essentially
exponentially. However, the relaxation time of $Z(t)$ increases with
increasing $T$. This counter-intuitive result can be
explained in terms of the ultrasoft nature of the potential: at higher
$T$ collisions become less and less effective in decorrelating the ion
velocities, since particles barely feel each other's influence. Thus,
the ``effective'' collision time  increases with increasing $T$. The
relaxation changes dramatically at the lower temperatures, where
$Z(t)$ decays much more slowly after an initially strong oscillatory
regime. This striking behavior may be rationalized in terms of
anion-cation pair formation. The oscillations correspond to the
vibrations of the ions relative to the CM of a ``bound'' pair. The
measured reduced angular frequency, $\omega \simeq 0.6$ is close to
the frequency of the relative motion of two oppositely charged ions
within the harmonic potential (see Eq.~(\ref{eq:v_origin}))
\begin{equation}
v_{+-}(r) = -u + u \frac{r^2}{12 \sigma^2}
\end{equation}
i.e., $\omega_{+-} = \sqrt{1/3} \simeq 0.58$, independent of
temperature. The slow decay of $Z(t)$ at long times may be associated
with the motion of the long-lived pair within which the ion is
bound. Due to the weakness of the interaction between pairs
(cf. Fig.~\ref{fig:effective_potentials_pp}), the motion of neutral
pairs is nearly ballistic, resulting in the very slow relaxation shown
in Fig.~\ref{fig:vacf}. The self-diffusion constant of the ions, $D_+
= D_- = D$ may be calculated by integrating the velocity
auto-correlation function (\ref{eq:vacf})~\cite{hansen_theory_2006}, or,
more accurately, from the asymptotic slope of the mean square
displacement of an ion from its initial position, according to
Einstein's relation~\cite{hansen_theory_2006}
\begin{equation}
D = \lim_{t \to \infty} 
\frac{\langle | {\bf r}(t) - {\bf r}(0) |^2 \rangle}{6t} .
\label{eq:self_diffusion}
\end{equation}
The corresponding ion mobility is $\mu = D/k_{\rm B}T$.

The variation of $D$ with temperature is shown in the inset of
Fig.~\ref{fig:vacf}. $D$ is seen to first drop as $T$ decreases, which
is the usual behavior observed in ``normal'' liquids before
increasing sharply for $T < 0.08$ and go through a pronounced maximum
around $T \simeq 0.03$ below which $D$ decreases again towards zero. This
unusual non-monotonic behavior is obviously a direct consequence of
pairing: as argued earlier pairing leads to a system of nearly
non-interacting entities which move almost freely, thus explaining
the sharp rise in mobility. Once pairing is complete and no free ions
are left, the diffusion constant will drop again with temperature.

\begin{figure}
\begin{center}
\includegraphics[width=\onefig,clip=true,draft=false,angle=0]
                {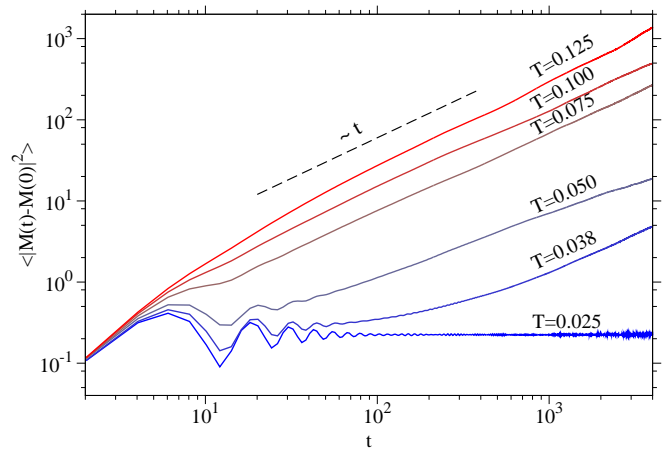}
\end{center}
\caption{Mean square displacement $\langle | {\bf M}(t) - {\bf M}(0)
  |^2 \rangle$ of the total electric dipole ${\bf M}(t)$ as a function
  of time at $n=0.0035$ for several temperatures (as labeled). The dashed line
  indicates the asymptotic behavior expected for a conducting system.}
\label{fig:m2msd}
\end{figure}

\begin{figure}
\begin{center}
\includegraphics[width=\onefig,clip=true,draft=false,angle=0]
                {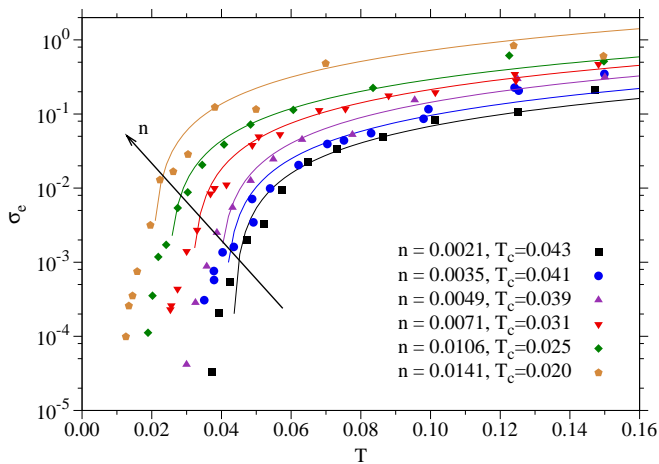}
\end{center}
\caption{Electrical conductivity $\sigma_e$, evaluated via
  Eq.~(\ref{eq:einstein}), as a function of temperature along
  different isochores (as labeled). $T_c$ indicates the estimated
  temperature where $\sigma_e$ vanishes, according to power-law fits
  $\sigma_e \sim (T - T_c)^\nu$ with $\nu=1.2$ (full lines). The
  estimated uncertainty on $\nu$ is $\pm 0.02$.}
\label{fig:conductivity}
\end{figure}

\begin{figure}
\begin{center}
\includegraphics[width=\onefig,clip=true,draft=false,angle=0]
                {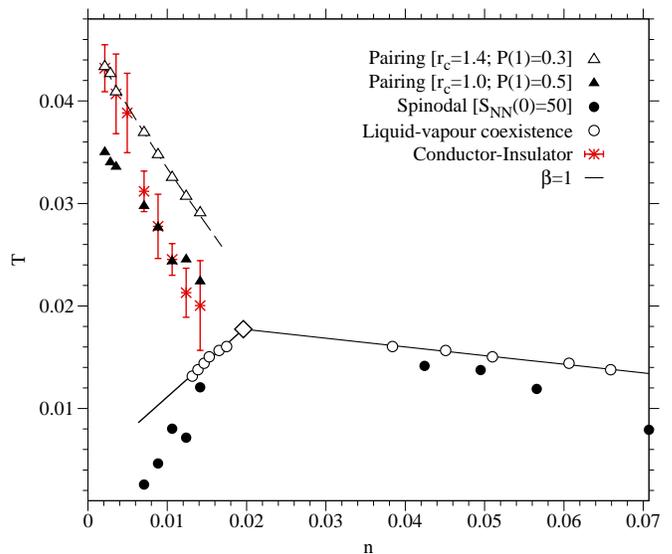} \\
\end{center}
\caption{Current estimate of the phase diagram of the URPM in the $(T, n)$-plane. Empty
  circles: coexistence points obtained via GCMC simulations (from
  Ref.~\cite{coslovich_clustering_2011}). Full line: fit to the
  coexistence line, assuming a critical exponent $\beta = 1$. Rotated empty
  square: estimate of the critical point assuming a critical exponent
  $\beta=1$. Filled circles: spinodal line estimated from the
  condition $S_{NN}(k = 0) \approx 50$. Open triangles: pairing transition
  points estimated from the condition $P(m=1)=0.3$ with $r_c=1.4$ (as
  in Ref.~\cite{coslovich_clustering_2011}). Filled triangles: pairing
  transition points estimated from the condition $P(m=1)\approx P(m=2)=0.5$
  with $r_c=1.0$. Stars: CI transition points estimated from the vanishing of the electrical conductivity (see text for definition).}
\label{fig:phase_diagram}
\end{figure}

We now turn to the electrical conductivity $\sigma_e$ per unit volume
of the URPM. $\sigma_e$ is determined by the asymptotic slope of the
mean square displacement of the total electric dipole ${\bf M}$,
according to the generalized Einstein relation
\begin{equation}
\sigma_e = \frac{1}{V k_{\rm B} T} \lim_{t \to \infty} 
\frac{ \langle | {\bf M}(t) - {\bf M}(0) |^2 \rangle}{6t}
\label{eq:einstein}
\end{equation}
which is equivalent to the familiar Green-Kubo relation linking
$\sigma_e$ to the electric current auto-correlation function
\cite{hansen_theory_2006}; in view of the large statistical uncertainties of 
the latter, as generated in MD
simulations, the Einstein relation (\ref{eq:einstein}) is preferable
to estimate $\sigma_e$. Data for the diffusion $c(t) = \langle | {\bf
  M}(t) - {\bf M}(0) |^2 \rangle$ of the total dipole along the isochore $n=0.0035$ are shown in
Fig.~\ref{fig:m2msd}. Note that ions must be allowed to leak out of
the periodically repeated simulations cell when calculating ${\bf
  M}(t)$, to allow for diffusion in an unbounded volume. The long-time
slopes of the $c(t)$-curves yield $\sigma_e$. Figure~\ref{fig:m2msd}
shows that the asymptotic linear regime is rapidly reached at the
higher temperatures, while at the lower temperatures the regime is
only reached after reduced times $t \gtrsim 10^3$. At the lowest temperature
investigated, $c(t)$ appears to tend to a constant, i.e., its slope and
hence $\sigma_e$, are zero, corresponding to an insulating state. Note
that $c(t)$ exhibits oscillations at the lower temperatures
reminiscent of those observed in $Z(t)$ (cf. Fig.~\ref{fig:vacf}),
of comparable frequency, which are once again ascribable to
pairing. 
We have repeated the electrical conductivity analysis along several
low density isochores. The data for the temperature dependence of
$\sigma_e$ are shown in Fig.~\ref{fig:conductivity}. The conductivity
is seen to drop to zero at increasingly low temperatures as the
density increases. Assuming a power-law dependence $\sigma_e \sim (T -
T_c)^\nu$ with a fixed value of $\nu \simeq 1.2$, least squares fits 
provide a set of ``critical'' temperatures $T_c$ at which $\sigma_e(T)$ appear to vanish (indicated by the labels 
in Fig.~\ref{fig:conductivity}). These results provide an estimate of a CI
transition line in the $(n, T)$ plane. This line is compared in
Fig.~\ref{fig:phase_diagram} to the pairing transition line which
provides another estimate of the location of the CI transition. The
two lines are seen to be roughly parallel and reasonably close. We
also include in Fig.~\ref{fig:phase_diagram} the pairing transition
line reported in Ref.~\cite{coslovich_clustering_2011}, in which
different values of $r_c$ and of the ``critical'' fraction of free
ions $P(m=1)$ were used. As suggested by Fig.~\ref{fig:permittivity},
different choices of $P(m=1)$ would correspond to different values of
a ``critical'' dielectric permittivity $\epsilon$.

\section{Discussion and conclusions}
\label{sec:conclusion_perspectives}

We have introduced a model of interpenetrating polyanions and
polycations, carrying extended, continuous charge distributions. We
expect this ultrasoft primitive model (UPM) to be relevant for the
study of the aggregation of oppositely charged polyelectrolytes in
good solvent. Most of the results presented in this paper are
restricted to the symmetric version of the model, the
URPM. Interactions between polyions are purely Coulombic; in
particular, contrary to the familiar restricted primitive model of
electrolytes, no hard cores are involved. We have reported extensive
simulation results and theoretical predictions characterizing the pair
structure, clustering, thermodynamic and dynamical properties of the
URPM over a wide range of temperatures and densities. Concerning the
static (structural and thermodynamic) properties, the RPA proves to be
a quantitatively reliable approximation at high densities,
corresponding to the fully ionic, conducting regime. At low
temperature and density, however, the RPA and even the HNC theory fail
completely, due to the formation of long-lived anion/cation pairs, and
some larger, mostly neutral clusters.

The low density, low temperature data for pairing, dielectric
permittivity $\epsilon$ and electrical conductivity $\sigma_e$ provide
strong evidence of a transition between an insulating phase,
characterized by finite values of $\epsilon$ and a vanishing
$\sigma_e$ (signaling a vanishing concentration of unpaired ions),
and a fully ionic, conducting state at higher $n$ and $T$, where
$\sigma_e$ takes on non-zero values and $\epsilon$ is very
large. Further, indirect evidence for a CI transition is provided by
the unusual variation of the polyion mobility with temperature at low
density.

At sufficiently low $T$ the CI transition gives way to a first order
phase separation between a low density insulating phase and a high
density conducting phase. The analysis of simulation data for
relatively small system sizes provides some preliminary evidence for the
existence of an upper tricritical point terminating the phase
coexistence line near the junction with the CI
line~\cite{coslovich_clustering_2011}. The exact nature of the CI
transition above the putative tricritical point is not entirely clear. The
simulation results presented herein and in Ref.~\cite{coslovich_clustering_2011}
point to a continuous transition, possibly broadened by finite size
effects. The tentative phase diagram seem to differ qualitatively from that
of the RPM, where criticality belongs almost certainly to the Ising
universality class
\cite{luijten_universality_2002,caillol_critical_2002}, and there is
no strong evidence for a CI line above $T_c$. The qualitatively
different behaviors of the hard core and penetrable electrolytes are
most probably linked to the very different nature of the anion/cation
pairs which dominate the insulating phase: they are weakly
interacting, polarizable entities in the case of the URPM, while they
are strongly interacting dipolar ``molecules'' in the
RPM~\cite{romero-enrique_dipolar_2002}. Interestingly, the phase
diagram of the URPM is more reminiscent of that of the
two-dimensional Coulomb gas involving logarithmic interactions between
ions (oppositely charged hard disks). In the low density limit, this
model is known to undergo an infinite order, Kosterlitz-Thouless
(KT)~\cite{kosterlitz_1973} transition between a dielectric phase of
bound ion pairs and a conducting phase of free ions. In the zero
density limit the KT transition is characterized by a discontinuous
jump of the dielectric permittivity from a finite value to infinity as
the transition temperature is approached from below. MC simulations on
finite systems show that the KT (or CI) transition is continuous
(rounded), and that the transition temperature drops with increasing
density; the CI line terminates close to the critical point of a first order vapor-liquid
coexistence curve~\cite{caillol_1986,orkoulas_1996}. Within statistical uncertainties, the simulations
show some evidence of a cusp at the top of the vapor-liquid
coexistence curve, suggesting a tricritical point (rather than the
regular critical point observed in the three-dimensional equivalent of
the Coulomb gas, namely the RPM). These findings are confirmed by mean
field calculations within the chemical representation of a mixture
of free ions and bound (Bjerrum) pairs~\cite{levin_1994}.

In order to confirm a similar scenario in the three-dimensional URPM,
it will be crucial to investigate finite size effects within a full
finite size scaling analysis, requiring extensive further simulations
for several system sizes (for a pedagogical review of finite size
scaling techniques, see~\cite{wilding_2001}). The natural order
parameters for such an analysis are $\Delta n$, i.e., the difference in density of the
two coexisting phases, and the fluctuation of the total dipole moment
per unit volume, which is intimately related to the dielectric
permittivity (see Eq.~(\ref{eq:permittivity})): such an investigation
is under way and will be the main object of a forthcoming publication.

Future developments of our work include an extension of the URPM to
non-symmetric versions of the UPM, which will involve less
straightforward aggregation patterns, and the generalization of the
model to oppositely charged polyions in the presence of microscopic
co- and counterions (e.g., from added salt) which will lead to screened
effective interactions between the polyions as opposed to the bare
Coulombic interactions considered in the present
work. The objective of this generalization will be to investigate the
influence of screening by microions on the CI transition and phase
separation.

\begin{acknowledgements} 
D.C. and G.K. acknowledge financial support by the Austrian Science
Foundation (FWF) under Project No. P19890-N16.
\end{acknowledgements}

\appendix

\section{Simulation methods}
\label{sec:appendix_a}

Numerical simulations of the URPM have been performed for systems
composed of $N=1000$ polyions in a cubic cell with periodic boundary
conditions using the Ewald summation scheme to
account for the long range of the interactions. In the following, we
briefly summarize the key equations and the parameters employed in our
simulations. Following a standard procedure~\cite{frenkel_smit_2001},
we introduce screening charge distributions of shape
\begin{equation}
\rho_\alpha^\prime(\ve{r}) = 
\left(\frac{1}{2\pi\sigma^\prime}\right)^{3/2}\exp{\left[-\frac{r^2}{2{\sigma^\prime}^2}\right]}
\end{equation}
such that the screening charge distribution around a particle of
species $\alpha$ is $-Z_\alpha \rho^\prime(\ve{r}-\ve{r}_i)$
[cf. Eq.~\eqref{eq:charge_density}]. Note that, in general, the width
$\sigma^\prime$ of the screening distribution needs not coincide with
the one of actual Gaussian distribution. Using the standard Ewald
summation scheme, the total interaction energy $U$ can then be
expressed as
\begin{equation}
  U = U_k + U_r - U_s
\end{equation}
where $U_k$ and $U_r$ are Fourier and real space contributions,
respectively, and $U_s$ is a self-term correcting for the interaction
between the screening distribution and its equal and opposite
compensating distribution. For the URPM, the three terms read
\begin{equation}
  \label{eq:ewaldurpm}
  U_k = \frac{1}{2V}\sum_{\{\ve{k}\}} \frac{4\pi}{k^2}
  \exp\left[-\frac{k^2 \tilde{\sigma}^2}{2} \right] |\rho_\ve{k}|^2
\end{equation}
\begin{equation}
  \label{eq:ewaldurpm1}
  U_r = \frac{1}{2} \sum_{i=1}^N {\sum_{i=1}^N}^\prime
  \frac{Z_i Z_j}{\epsilon} \left[ \frac{\text{erf}(r/2\sigma)}{r}
                               - \frac{\text{erf}(r/\sqrt{2}\tilde{\sigma})}{r} 
                             \right]
\end{equation}%
\begin{equation}
  \label{eq:self}
  U_s = \frac{1}{\sqrt{2\pi}\tilde{\sigma}} \sum_{i=1}^{N} Z_i^2
\end{equation}
where $\rho_\ve{k} = \sum_i Z_i \exp{(i\ve{k}\cdot\ve{r}_i)}$ and we
have introduced the notation
$\tilde{\sigma}=\sqrt{\sigma^2+{\sigma^\prime}^2}$. Note that the
familiar expression for $U_k$ in the case of purely Coulombic
interaction is recovered by substituting $\tilde{\sigma}\rightarrow
\sigma$ in Eq.~\eqref{eq:ewaldurpm}. Expressions for the forces
between charge distributions (needed in MD
simulations) are derived in a similar manner.

When $\sigma^\prime = \sigma$ the real space contribution is
identically zero and the calculation of the potential energy and the
forces can be carried out entirely in Fourier space. Note that this is
possible only because the effective potential $v_{\alpha\beta}(r)$
[see Eq.~\eqref{eq:potential}] is bounded. The sum in Eq.~\eqref{eq:ewaldurpm}
is carried out for all wave vectors $\ve{k}$ such that $|\ve{k}|$ is
smaller than some cut off value $k_c$. The error in the evaluation of
$U_k$ can then be roughly estimated as~\cite{smith_forester_1996}
\begin{equation}
  \label{eq:fourierspace} 
  \varepsilon = \frac{1}{\epsilon} \frac{\exp{\left(-k_c^2\sigma^2\right)}}{k_c^2} .
\end{equation}

\begin{figure}
\begin{center}
\includegraphics[width=\onefig,clip=true,draft=false,angle=0]
                {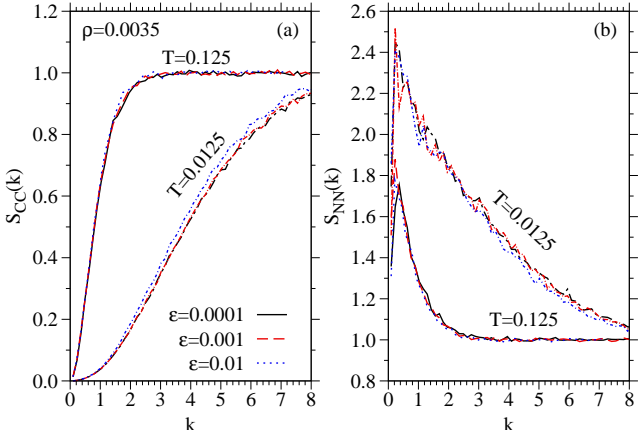} \\
\end{center}
\caption{(a) Concentration-concentration structure factors $S_{CC}(k)$
  and (b) number-number structure factors $S_{NN}(k)$ at $n=0.0035$ and
  different temperatures (as labeled), obtained using different
  values of the Ewald sum precision $\varepsilon$:
  $\varepsilon=10^{-4}$ (full lines), $\varepsilon=10^{-3}$ (dashed
  lines), and $\varepsilon=10^{-2}$ (dotted lines).  }
\label{fig:precision}
\end{figure}

In our simulations, we used indeed $\sigma^\prime = \sigma$ and
adjusted $k_c$ so as to keep the nominal error $\varepsilon$ constant
at $10^{-3}$. As a consequence, the actual number $N_k$ of
wave-vectors used for the evaluation of $U_k$ varied as a function of
the density (as $N_k\sim V^{-1/3}$), ranging from $2.5 \times 10^2$ at
$n=0.35$ to $4.2 \times 10^4$ at $n=0.0035$. We found that the chosen
value of $\varepsilon$ was sufficient to converge structural and dynamic
properties below the noise level. To illustrate this, in
figure~\ref{fig:precision} we compare results for the structure
factors $S_{NN}(k)$ and $S_{CC}(k)$ at $n=0.0035$ obtained
using $\varepsilon=10^{-2}$, $10^{-3}$, and $10^{-4}$. We note that
the effect of reducing $\varepsilon$ becomes slightly more pronounced
at low $T$. We remark that these small discrepancies are, however, 
irrelevant for the purposes of this work.

At even lower densities ($n\alt 0.001$), the number of wave-vectors
required to keep $\varepsilon$ constant becomes too large for an
efficient computation of Eq.~\eqref{eq:ewaldurpm}. In this regime, a simple
decimation strategy could be used to reduce the number of wave-vectors
in each spherical shell $k\pm \delta k$. Alternatively, one could use
screening charge distributions with width $\sigma^\prime\ne \sigma$,
so as to move part of the computational effort into real space. In
practice, however, we hardly obtained any benefit from using an
optimized $\sigma^\prime$, at least in the range of state parameters
investigated in the current study. In must be noted, in fact, that the
range of the pair potential entering Eq.~\eqref{eq:ewaldurpm1} is
relatively long compared to the complementary error function, normally
encountered for pure Coulombic interactions. 
As a
consequence, a large cut off must be employed to evaluate the real
space contribution $U_r$. Therefore, complete splitting of the
interaction into real space and Fourier space contributions does not
appear particularly convenient for the system at hand.

MD simulations were carried out in the microcanonical
ensemble (NVE) using the velocity Verlet algorithm with a time step
$\delta t = 0.08$. Equilibration at the temperature of interest was
achieved by coupling the system to a massively stochastic heat
bath~\cite{allen_tildesley_1987}. Monte Carlo simulations have been
performed in the canonical (NVT) and grand-canonical 
($\mu$VT) ensemble. To accelerate equilibration at low density and temperature,
we also implemented cluster moves to displace and insert/delete pairs
of ions, as described in Ref.~\cite{orkoulas_free_1994}. In NVT
simulations, the maximal displacements of standard displacement moves
and cluster moves were adjusted during equilibration so as to keep the
acceptance ratio around 30\%.  In $\mu$VT simulations, both random and
biased insertions/deletions of pairs were attempted, together with
standard displacement and cluster moves. Analysis of the acceptance
ratios of Monte Carlo moves as a function of state parameters showed
that biased insertions/deletions become favorable close to the
estimated critical region, as expected. Proper equilibration of the
system at low density was checked by comparing results of different
thermal histories and the different simulation methods. We remark that
equilibration may become a serious issue at densities and temperatures
lower than the ones considered in this work. In that case, more
efficient simulation methods should be
used~\cite{valeriani_computer_2010}.

\section{Effective interactions}
\label{sec:appendix_b}

In this Appendix a brief derivation is provided of the effective
interaction between two anion/cation pairs, and between a free ion and
a pair, in the limit of large separations. Consider first the case of
two pairs. Let ${\bf r}_1$ and ${\bf r}_2$ denote the distances
between the two oppositely charged polyions in the two pairs, and let
${\bf R}$ be the vector joining the CM's of the pairs. For
sufficiently low temperatures $T$, the classical relative vibrations
of the two ions in a pair are of small amplitude, so that their
interaction potential $v_{+-}(r)$ may be replaced by its small $r$
limit (\ref{eq:v_origin}), i.e., $\beta v_{+-}(r) = - \beta u +
\frac{\gamma}{2} r^2$ with $u = Q^2/(\sqrt{\pi} \epsilon ' \sigma)$ and $\gamma
= \beta u/6 \sigma^2$. The configurational part of the internal
partition function of an isolated pair is hence
\begin{equation}
q_1 = {\rm e}^{\beta u} \int {\rm e}^{-\gamma r^2/2} d {\bf r} =
{\rm e}^{\beta u} \frac{(2 \pi)^{3/2}}{\gamma^{3/2}} .
\end{equation}
The configurational part of the partition function of two interacting
pairs separated by ${\bf R}$ is
\begin{eqnarray}\nonumber
&q_2(R) = \int d {\bf r}_1 \int d {\bf r}_2 
\exp \left \{- \beta \left[ v_{+-}(r_1) \right.\right.\\
&\left.\left.+ v_{+-}(r_2)
+ v_{12}({\bf R}, {\bf r}_1, {\bf r}_2) \right] \right\} .
\label{eq:q2}
\end{eqnarray}
If $R \gg \sigma$ one may adopt the point dipole approximation for the
pair/pair interaction, each carrying a dipole ${\bf m}_i = Q {\bf
  r}_i$
\begin{equation}
\beta v_{12}({\bf R}, {\bf r}_1, {\bf r}_2) =
\frac{\beta Q^2}{\epsilon'} \left[ \frac{ {\bf r}_1 \cdot {\bf r}_2}{R^3}
- \frac{3 \left( {\bf R} \cdot {\bf r}_1 \right) \left( {\bf R} \cdot {\bf r}_2 \right)}
{R^5} \right] .
\label{eq:dipole_approximation}
\end{equation} 
Substituting (\ref{eq:dipole_approximation}) into (\ref{eq:q2}), and
choosing ${\bf R}$ to be the polar axis, $q_2$ may be cast in the form
\begin{equation}
  \begin{split}
& q_2(R) = 2 \pi {\rm e}^{2 \beta u} 
\int_0^\infty\!\! {\rm e}^{-\gamma r_1^2/2} r_1^2 d r_1 
\int_0^\infty\!\! {\rm e}^{-\gamma r_2^2/2} r_2^2 d r_2 \\ 
 &\int_0^{2 \pi}\!\! d \varphi \int_0^\pi\!\! \sin \Theta_1 d \Theta_1 
\int_0^\pi\!\! \sin \Theta_2 d \Theta_2 \\
& \exp \left \{ - \frac{\beta Q^2}{\epsilon '} \frac{r_1 r_2}{R^3} 
\left[ \sin \Theta_1 \sin \Theta_2 \cos \varphi - 2 \cos \Theta_1 \cos \Theta_2 \right] \right \}
  \end{split}
\label{eq:q2_explicit}
\end{equation}
where $\varphi = \varphi_1 - \varphi_2$. For sufficiently large $R$
the last exponential in the integrand of~\eqref{eq:q2_explicit} may
be Taylor expanded. Elementary calculations show that all odd power
terms in the expansion vanish upon carrying out the angular
integrations. Retaining the zeroth and second order terms in the
Taylor expansion one arrives at
\begin{equation}
q_2(R) = {\rm e}^{2 \beta u} \frac{(2 \pi)^3}{\gamma^3} 
\left[ 1 + 108 \pi \left( \frac{\sigma}{R} \right)^6 \right] .
\end{equation}

The effective pair-pair potential is then given by the free energy of
the interacting pairs minus the sum of the internal free energies of
each pair, i.e.,
\begin{eqnarray}\nonumber
\beta v_{2,2}(R) = \beta f_2(R) = - \ln \left[ \frac{q_2(R)}{(q_1)^2} \right]
= \\- \ln \left[1 + 108 \pi \left( \frac{\sigma}{R} \right)^6 \right]
\simeq - 108 \pi \left( \frac{\sigma}{R} \right)^6
\label{eq:v_22}
\end{eqnarray}
valid for $R \gg \sigma$. The ``van der Waals'' exponent 6 agrees with
the asymptotic form of the effective potential extracted from MC
simulations (Fig.~\ref{fig:effective_potentials_pp}). The next
contribution would be ${\cal O}(1/R^{12})$.

The effective ion/pair potential $v_{1,2}(R)$ may be calculated along
similar lines, starting from the coupling between a single ion and a pair
of instantaneous dipole moment $Q {\bf r}$
\begin{equation}
\beta v({\bf R}, {\bf r}) = \frac{\beta Q^2}{\epsilon '} 
{\bf R} \cdot {\bf r} .
\end{equation}
The resulting effective pair potential is found to be
\begin{equation}
\beta v_{1,2}(R) = - \frac{3 \pi}{T} \left( \frac{\sigma}{R} \right)^4 .
\end{equation}
Note that, contrary to $\beta v_{2,2}(R)$ in Eq.~(\ref{eq:v_22}) $\beta v_{1,2}(R)$
depends on temperature. The same calculation leads to the following
expression for the reduced polarizability of an anion/cation pair
\begin{equation}
\frac{\xi}{\sigma^3} = 6 \sqrt{\pi} \epsilon ' .
\end{equation}

%


\end{document}